\documentclass{mn2e}
\usepackage{epsfig}
\usepackage{rotating}
\usepackage{longtable}
\usepackage{graphics}
\usepackage{mathrsfs}
\usepackage{amssymb}
\usepackage{supertabular}
\usepackage{times}

\newcommand{\apj}{ApJ}

\newcommand{\aap}{A\&A}
\newcommand{\aj}{AJ}
\newcommand{\apjl}{ApJL}
\newcommand{\mnras}{MNRAS}
\newcommand{\araa}{ARA\&A}
\newcommand{\hcop}{HCO$^+$}
\newcommand{\nthp}{N$_2$H$^+$}

\newcommand{\msol}{M$_\odot$}
\newcommand{\td}{$T_d$($r$)}
\newcommand{\tg}{$T_k$($r$)}
\newcommand{\tex}{$T_{\rm exc}$}
\newcommand{\tp}{$T_{\rm PPC}$}
\newcommand{\tc}{$T_{\rm cls0}$}
\newcommand{\tcol}{$t_{\rm coll}$}
\newcommand{\ncr}{$n_{\rm cr}$}
\newcommand{\cmt}{cm$^{-3}$}

\title[Modelling the `blue asymmetry' in inside-out collapse]
{Molecular line profiles as diagnostics of protostellar collapse: modelling the
`blue asymmetry' in inside-out infall}

\author[Y. G. Tsamis et al.]
{Y.~G. Tsamis, J. M. C. Rawlings, J. A. Yates, and S. Viti\\
Department of Physics and Astronomy, University College London, Gower Street,
London WC1E 6BT, U.K.; E-mail: ygt@star.ucl.ac.uk}

\date{Accepted 2008 May 12}

\begin{document}
\maketitle

\begin{abstract}

\noindent The evolution of star-forming core analogues undergoing inside-out
collapse is studied with a multi-point chemodynamical model which
self-consistently computes the abundance distribution of chemical species in
the core. For several collapse periods the output chemistry of infall tracer
species such as \hcop, CS, and \nthp, is then coupled to an accelerated
$\Lambda$-iteration radiative transfer code, which predicts the emerging
molecular line profiles using two different input gas/dust temperature
distributions. We investigate the sensitivity of the predicted spectral line
profiles and line asymmetry ratios to the core temperature distribution, the
time-dependent model chemistry, as well as to {\sl ad hoc} abundance
distributions. The line asymmetry is found to be strongly dependent on the
adopted chemical abundance distribution. In general, models with a warm central
region show higher values of blue asymmetry in optically thick \hcop\ and CS
lines than models with a starless core temperature profile. We find that in the
formal context of Shu-type inside-out infall, and in the absence of rotation or
outflows, the \emph{relative} blue asymmetry of certain \hcop\ and CS
transitions is a function of time and, subject to the foregoing caveats, can
act as a collapse chronometer. The sensitivity of simulated \hcop\ line
profiles to linear radial variations, subsonic or supersonic, of the internal
turbulence field is investigated in the separate case of static cores.

\vspace{0.3cm}


\end{abstract}

\section{Introduction}

The study of star-formation is entering a new era with the advent of large
(sub-)millimetre arrays such as the eSMA (SMA-JCMT-CSO combined), CARMA, and
later ALMA. These facilities will provide crucial new information on the gas
kinematics in pre-stellar cores, proto-stars and circumstellar disks in nearby
star-forming clouds, matching, or in the case of ALMA, exceeding the spatial
resolution of the {\it Hubble Space Telescope}. The {\it Spitzer Space
Telescope} has recently mapped the infrared emissions from a large number of
candidate, low-mass young stellar objects belonging to Classes 0, I, II in
nearby ($\sim$\,200--300~pc) interstellar clouds (e.g. J$\o$rgensen et al.
2006; Brooke et al. 2007; Porras et al. 2007). No doubt these newly detected
sources will later become ideal targets for the large arrays.

Of particular interest when studying the earliest stages of star-formation, as
in starless `pre-protostellar' cores (PPCs; Ward-Thompson et al. 1994), and
Class~0 sources, is the identification and characterization of the associated
gas infall, which requires the kinematic interpretation of the observed
molecular line profiles. This is by no means straightforward and in some cases
the approach has been based on statistics: key core-collapse indicators, such
as the relative number of blue- versus red-shifted line profiles (quantified as
a normalized velocity difference), or the `blue' to `red' peak intensity ratio,
are analyzed drawing from a bulk sample of infall candidates (e.g. Mardones et
al. 1997; Gregersen et al. 2000; Gregersen \& Evans 2000; Sohn et al. 2007).
These studies can be greatly assisted by detailed modelling of pre-stellar
cores; such models, including sophisticated molecular line radiative transfer
(RT) and/or dust RT calculations, have been presented by Evans et al. (2001),
Ward-Thompson \& Buckley (2001), Redman et al. (2004), De Vries \& Myers
(2005), and Pavlyuchencov et al. (2006) amongst others. The studies of Redman
et al. and Pavlyuchenkov et al., in particular, have demonstrated how effects
other than infall, such as rotation of the core, can be inferred from a proper
modelling of the observed line profiles.

The large number of poorly constrained free parameters in the models suggests
two alternative approaches to the problem; (i) generic modelling, where the
sensitivities of the line profiles to the various free parameters are analyzed
-- with a view to identifying and breaking possible degeneracies, and (ii)
modelling individual sources/datasets, making plausible assumptions concerning
the physical and chemical natures of the source. This study adopts the former
approach and follows the exploratory work of Rawlings \& Yates (2001; hereafter
RY01). RY01 presented preliminary results obtained from coupled
chemical/dynamical and line RT models applicable to generic, low-mass
(1--3\,\msol) infalling cores. In that paper they explored the sensitivity of
the emerging line profiles to limited variations in a subset of the free
parameters in the chemistry and dynamics. Although they only considered two
transitions (HCO$^{+}$~$J$ $=$ 4$\to$3 and CS $J$ $=$ 3$\to$2), they found that
the line profiles are {\em extremely} sensitive to assumed gas-grain
interaction efficiencies and the dynamical history of the core. The line
profiles are less sensitive to the cosmic ray ionization rate (as a consequence
of the fact that the timescale for the gas-phase chemistry is less significant
than the freeze-out timescale).

As an example of the second approach, Evans et al. (2005) attempted to model 25
transitions of 17 isotopologues in the well-known infall source B335 (e.g. Zhou
et al. 1993). The model included a careful consideration of the thermal balance
and calculated both the gas and dust temperatures. Assuming that the dynamics
of the source can be described by the Shu (1977) `inside-out' collapse model,
preceded by the quasi-static contraction of a sequence of Bonnor-Ebert spheres,
reasonably good matches with the observational data can be obtained, provided
the cosmic ray ionization rate and the sulphur abundance are enhanced. However,
it must be emphasized, that this result whilst {\em compatible} with the
assumed chemistry and dynamics does not not unequivocally {\em validate} those
assumptions. The standard model of low-mass star formation which advocates that
a core collapses via ambipolar diffusion, followed by inside-out infall after
it has become supercritical (Shu et al. 1987), has been challenged by
observations showing evidence of extended inward motions and non-zero infall
velocities ($\sim$0.1 km\,s$^{-1}$) at large core radii ($\sim$10$^4$ AU); see,
for instance, sources such as L1544 (Tafalla et al. 1998), IRAM 04191 (Belloche
et al. 2002), L1551 (Swift et al. 2006). Observations of B335, on the other
hand, have been adequately explained by the standard model (Zhou et al. 1993;
Choi et al. 1995). Even for this most well studied infall source, however, the
exact interpretation of the blue/red line asymmetry for some molecular tracers
is further complicated by the presence of an outflow (e.g. Choi 2007).

An analysis of 50 PPCs  by Sohn et al. (2007) estimated the amount, $\delta V$,
by which the HCN ($J$ $=$ 1--0) hyperfine spectrum is blue- or red-shifted with
respect to the line centre of the optically thin tracer \nthp\ ($J$ $=$ 1--0),
and revealed that the $\delta V$ distribution of the HCN ($J$ $=$ 0--1)
components is skewed to the blue consistent with systematic inward motions in
the cores -- the degree of skewness was greater than for the other infall
tracers they used (CS $J$ $=$ 2--1, DCO$^+$ $J$ $=$ 2--1, and \nthp\ $J$ $=$
1--0). In four out of 12 infall candidates (L63, L492, L694-2, and L1197) the
blue to red peak asymmetry ratio was larger for the hyperfine line with the
lowest opacity ($F$ $=$ 0--1), which is a better tracer of dense inner regions,
suggesting that these cores may show \emph{increasing} gas velocities towards
their centres. The exact shape of the velocity profiles in starless cores,
however, is an open issue: for example, Lee et al. (2007) have modelled two of
these cores adopting a `$\Lambda$-shaped' profile which calls for an
approximately zero-infall velocity in the innermost core regions (as opposed to
inside-out collapse where the velocity there is the largest) with a peak in
velocity somewhere in the intermediate regions. This contradicts the work of
Williams, Lee and Myers (2006) who modelled one of the cores in the Lee et al.
study with a velocity profile increasing towards the centre. Clearly, the
models may not be unique and the adopted velocity laws may be representative of
only a particular evolutionary phase of a core.

In this study we adhere to the inside-out model of collapse and extend the
analysis of RY01 concentrating on the effects of the thermal structure and the
molecular abundance of infall tracer species on the emerging line asymmetry (in
the form of the peak blue/red ratio -- not the normalized velocity difference
of Mardones et al. 1997). We adopt models of low-mass ($\sim$3\,M$_\odot$)
static and infalling cores subject to time-dependent chemical evolution, and
investigate the various line profiles simulated by means of an accelerated
$\Lambda$-iteration (ALI) RT code. The main focus of this paper is a
theoretical analysis of the line asymmetry in the context of the inside-out
paradigm of protostellar collapse. The hydrodynamical, chemical and RT models,
and model assumptions, are presented in Section 2. The results are presented in
Section 3, and our conclusions in Section 4.

\section{The Model}

As described above when trying to analyze the molecular line profiles, there
are a number of free-parameters that need to be defined/constrained. These can
be summarized as follows:

\noindent (i) The instantaneous density and temperature profiles, $n(r)$, and
$T(r)$.

\noindent (ii) The instantaneous systemic velocity profile, $v(r)$.

\noindent (iii) The instantaneous microturbulent velocity profile, $\sigma(r)$.

\noindent (iv) The chemical abundance profiles, $n_{\rm mol}(r)$. \\

\noindent The chemical abundance profiles are particularly hard to constrain as
they can (critically) depend on the assumed initial conditions and the {\em
history} of the object. For example, Rawlings et al. (2002) showed that the
chemical evolution of many species is highly dependent on the assumed initial
H:H$_2$ ratio. Moreover, as the dynamical and chemical kinetic timescales are
comparable ($\sim$ 10$^6$\,yr), the abundance profiles are sensitive to the
dynamical evolution in both the collapse {\em and} the pre-collapse phases. As
well as the dependencies on the cosmic ray ionization rate and the elemental
abundances etc., the chemistry is critically dependent on the assumptions
concerning the gas-grain interactions: e.g. the effective grain surface areas,
the sticking efficiencies, the desorption mechanisms (and efficiencies) and the
surface chemistry.

\subsection{The hydrodynamics and chemistry}

\begin{figure}
\centering \epsfig{file=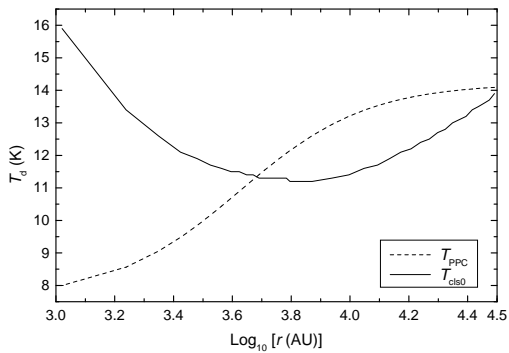, width=5.5 cm, scale=, clip=, angle=-90}
\caption{The two kinetic temperature distributions in our inside-out collapsing
core models.}
\end{figure}

Each of the free parameters discussed above should be considered in a full
sensitivity analysis. In this study we concentrate on a subset. This is partly
for the sake of brevity and clarity, and partly because we consider these free
parameters to be particularly important.

The density and dust temperature profiles [$n(r)$, $T_d(r)$] of protostellar
cores are now reasonably well-constrained by continuum data. For the dust
temperature, \td, we use radial profiles appropriate for PPCs and Class~0
sources: for the former case we adopted the generic profile computed by Evans
et al. (2001; their fig. 3) who used a 1D continuum radiative transfer model;
for the latter case the radial profile is the one deduced by Zhou et al. (1990)
from dust continuum observations of B335 for a luminosity of $\sim$
3\,L$_\odot$ (again using a radiative transfer model -- see also Shirley, Evans
\& Rawlings 2002 who obtained very similar temperature profiles). These two
profiles respectively denoted \tp\ and \tc\ are shown in Fig.\,~1. In both
cases we assume that the dust and kinetic gas, \tg, temperatures are thermally
coupled, i.e. \tg\ $=$ \td, an assumption which probably holds in the denser
line-forming regions of the core's center. Observationally, variations in the
temperature profiles within each protostellar class are relatively small, so we
make the simplification that the profiles do not vary with time. As the level
populations are only dependent on the instantaneous temperatures and the range
of temperatures involved ($\sim$ 8--16\,K) is too small to lead to chemical
variations, the history of the temperature profiles is of little significance
(see however Section~3.3).

Note that, although all of the results that we present in this work are for
inside-out collapsing cores, we have considered both the \tp\ and \tc\
temperature profiles. This allows us to determine the sensitivity of the
simulated line profiles to the assumed temperature distribution, even though
(i) a Shu-type collapse is not, in all probability, universally applicable to
protostellar infall sources, and (ii) realistically a \tp\ temperature profile
is {\em not} expected following the formation of the singular isothermal
spheroid (SIS) which forms the basis of the inside-out collapse.

In this study we do not investigate the sensitivities to the temporal evolution
of the density and velocity laws either during the collapse, or the
pre-collapse phases. Nor do we consider the possibility of the presence of
non-thermal (i.e. magnetic) support. Instead, we assume a purely gravitational
collapse, and impose the exact dynamical collapse solution of Shu (1977). For
the physical and chemical `pre-evolution', we follow RY01 and adopt the initial
conditions of an irradiated diffuse cloud, of density 10$^3$\,cm$^{-3}$, that
is in a state of chemical equilibrium, and allow it to contract to the SIS
configuration that forms the initial basis for the Shu (1977) model. This
contraction is represented by a retarded free-fall collapse parametrized as in
Rawlings et al. (1992) and has a duration of the order of 10$^6$ yr. The
extinction is dynamically calculated as a function of position and time, both
during this initial collapse phase, and also during the subsequent inside-out
collapse. This approach, although somewhat arbitrary, is probably reasonably
representative of the (empirically unconstrained) evolution of a cloud from a
diffuse state to a pre-collapse core. We defer a full analysis of the
sensitivities of the line profiles to the dynamics to a later study.

As emphasized above, gas-grain interactions are very poorly constrained -- a
fact that is often overlooked in studies of this type. For instance, Roberts et
al. (2007) have showed that the desorption mechanisms and efficiencies are
particularly uncertain. They advocate a more empirical approach, which is what
we adopt in this study: thus, the dust grains, which are characterized by a
population-averaged surface area (as in Rawlings et al. 1992), are treated only
as an inert `sink' for gas-phase species and the chemistry consists solely of
gas-phase chemical reactions. The freeze-out of gas-phase material onto the
surface of grains is represented with a nominal sticking efficiency of
$S_{i}=0.3$ for most species. We additionally assume that freeze-out only
occurs above some critical value of the extinction, $A_v(crit.)$. The value
that we adopt in this study, $A_v(crit.)$ $=$ 4.0, is similar to that deduced
for Taurus ($=$ 3.3) based on the observed correlations between the 3$\mu$m
water ice feature and visual extinction (Whittet et al. 1988).

\setcounter{table}{0}
\begin{table}
\caption{Parameters used in the dynamical/chemical models.}
\begin{center}
\begin{tabular}{|ll|}
\hline
Sticking co-efficient (S$_i$) & 0.3 \\
Cosmic ray ionization rate ($\zeta$) & $1.0\times10^{-16}$s$^{-1}$ \\
%Duration of hydrodynamic pause (t$_{pause}$) & 0 years \\
External extinction ($A_V$~ext.) & 3.0 mag \\
Critical extinction ($A_V$~crit.) & 4.0 mag \\
Cloud radius & 0.15\,pc \\
Effective sound speed & 0.21 km s$^{-1}$ \\
\noalign{\vskip2pt} \multicolumn{2}{c}{{\it Collapse period ($t_{\rm coll}$)
and infall radius
($R_{\rm CEW}$)}} \\
\noalign{\vskip2pt}
\multicolumn{2}{c}{($t_{\rm coll}$, $R_{\rm CEW}$) $=$ (1.36$\times$10$^5$\,yr,
0.03\,pc)} \\
\multicolumn{2}{c}{($t_{\rm coll}$, $R_{\rm CEW}$) $=$ (2.72$\times$10$^5$\,yr,
0.06\,pc)}\\
\multicolumn{2}{c}{($t_{\rm coll}$, $R_{\rm CEW}$) $=$ (4.08$\times$10$^5$\,yr,
0.09\,pc)}\\
\multicolumn{2}{c}{($t_{\rm coll}$, $R_{\rm CEW}$) $=$ (5.44$\times$10$^5$\,yr,
0.12\,pc)}\\
\hline
\end{tabular}
\end{center}
\end{table}

\setcounter{table}{1}
\begin{table}
\caption{Control parameters for {\sc smmol}. See text for details.}
\begin{center}
\begin{tabular}{|ll|}
\hline
Number of radial shells & 100 \\
Number of lines of sight & 500 \\
Number of frequency bins & 100 \\
Convergence criterion & 1$\times 10^{-4}$ \\
Distance to source & 500 pc \\
Telescope diameter & 45 m \\
%Beam efficiency & 0.6 \\
\multicolumn{2}{c}{{\it Free parameters}} \\
\noalign{\vskip2pt}
\td      &PPC or Class~0 profile (Fig. 1) \\
$n_{\rm mol}(r)$ & `model' or arbitrary (Fig. 2) \\
%\noalign{\vskip2pt}
\end{tabular}
\end{center}
\begin{center}
\begin{tabular}{|crrr|}
\hline
%\noalign{\vskip2pt}
%\multicolumn{4}{c}{Transitions, frequencies (GHz) and FWHM beam sizes ($''$)}
\\
Transition     & $\nu$ (GHz) & HPBW ($''$) &Beam effic. \\
\multicolumn{4}{c}{~~~~\hcop} \\
$J$ $=$ 1$\rightarrow$0 &89.19         & 18.8 &0.49\\
~~~~~~~~2$\rightarrow$1 &178.38        & 9.4  &0.63\\
~~~~~~~~3$\rightarrow$2 &267.56        & 6.3  &0.44\\
~~~~~~~~4$\rightarrow$3 &356.74        & 4.7  &0.62\\
~~~~~~~~5$\rightarrow$4 &445.90        & 3.8  &0.53\\
\multicolumn{4}{c}{~~~~CS}      \\
$J$ $=$ 1$\rightarrow$0 &48.99         & 34.2 & 0.59\\
~~~~~~~~2$\rightarrow$1 &97.98         & 17.1 & 0.49\\
~~~~~~~~3$\rightarrow$2 &146.97        & 11.4 & 0.69\\
~~~~~~~~4$\rightarrow$3 &195.95        & 8.6  & 0.60\\
~~~~~~~~5$\rightarrow$4 &244.94        & 6.8  & 0.49\\
\multicolumn{4}{c}{~~~~\nthp}      \\
$J$ $=$ 1$\rightarrow$0 &93.17         & 18.0 & 0.49\\
%~~~~~~~~2$\rightarrow$1 &186.34        & 9.0 & 0.62\\
%\multicolumn{4}{c}{~~~~p-\amm}      \\
%$J$ $=$ 1$\rightarrow$0 &23.69         & 70.8 & 0.71\\
%~~~~~~~~2$\rightarrow$1 &23.72         & 70.7 & 0.71\\
%\multicolumn{4}{c}{~~~~o-\amm}      \\
%$J$ $=$ 1$\rightarrow$0 &572.79         & 2.9 & 0.41\\
%~~~~~~~~2$\rightarrow$1 &1215.4         & 1.4 & 0.24\\
\hline
\end{tabular}
\end{center}
\end{table}

We have used a multi-point chemical model, similar to the one employed by RY01,
that follows the chemical and dynamical evolution of a grid of (50--100) points
in a Lagrangian co-ordinate system. We follow the time and space evolution of
89 gas-phase and 36 solid-state chemical species which contain the elements H,
He, C, N, O, S and Na, as well as electrons, through a comprehensive network of
some 2000 gas phase reactions and 167 freeze-out reactions. The chemistry is
limited to species containing four atoms or less. The total (depleted)
elemental abundances are, by number, relative to H; He\,:\,0.1,
C\,:\,1.88$\times10^{-4}$, N\,:\,1.15$\times10^{-4}$,
O\,:\,6.74$\times10^{-4}$, S\,:\,1.62$\times10^{-7}$, and
Na\,:\,3.5$\times10^{-7}$. These values include gas-phase depletion factors of
0.5, 0.01 and 0.01 for C, S and Na respectively relative to their cosmic
values. The chemical reaction network and rates have been drawn from the UMIST
rate-file databases (e.g. Millar, Farquhar \& Willacy 1997; Millar et al.
1991), and a full set of photo-reactions (including those induced by the cosmic
ray ionization of H$_2$) are included.

The physical parameters that we have fixed in this study are the {\it external}
extinction, $A_v(ext.)$, along the line of sight to the edge of the core, the
cosmic ray ionization rate ($\zeta$) and the effective sticking coefficients
($S_i$). We assume that all species have the same basic sticking coefficient.
Variables include: (i) the thermal structure of the core, designated \tp\ or
\tc\, as described above (Fig.\,~1), and (ii) the abundance distribution of
chemical species in the core. To identify the sensitivities we have considered
four possible abundance distributions, whose
generic form is shown in Fig.\,~2:\\

\noindent (i) As given by the chemodynamical model. These abundance profiles
vary during the evolution of the collapse in a manner similar to that shown in
RY1 (cf. their fig.\,~4ab).

\noindent (ii) Monotonically increasing with radius (arbitrary but within the
bounds of the chemical model).

\noindent (iii) Monotonically decreasing with radius (arbitrary as above).

\noindent (iv) Constant throughout the core. For these models we use fractional
abundances for HCO$^+$ and CS (relative to molecular hydrogen) of a few
$\times$ $10^{-8}$, which is within the range of values obtained by the
chemical model.\\

\noindent For the collapse dynamics we use values that are close to those
deduced in the Zhou et al. (1993) model, but consider various collapse periods
($t_{\rm coll}$; the duration since the initiation of inside-out collapse)
along with their corresponding infall radii ($R_{\rm CEW}$). All these are
summarized in Table 1. The form of the systemic velocity law is similar to the
one shown by RY01 in their fig.\,~1b, and is assumed to be the same for all
models. The value for the microturbulent velocity ($\sigma_{\rm NT}$ $\sim$
0.145 km\,s$^{-1}$) was also taken from Zhou et al. and, in the standard model,
is taken to be the same at all positions and times.

%The effective sound speed in the cloud is taken to be 0.21 km s$^{-1}$, as is
%appropriate for cold molecular gas.

\subsection{Radiative Transfer}

\begin{figure}
\centering \epsfig{file=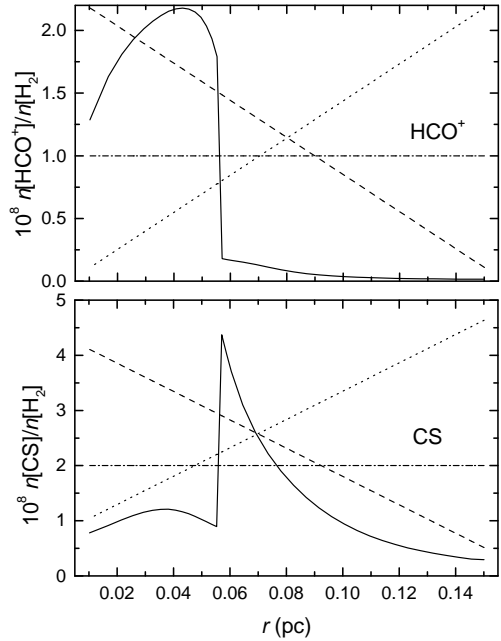, width=7.5 cm, scale=, clip=, angle=0}
\caption{The linear fractional abundances of \hcop\ ({\it top}) and CS ({\it
bottom}) as a function of the core radius at $t_{\rm coll}$ $=$
2.72$\times$10$^5$\,yr. The various curves correspond to the `model' ({\it
solid}), `constant' ({\it dot-dash}), `increasing' ({\it dot}), and
`decreasing' ({\it dash}) abundance distributions discussed in the text.}
\end{figure}

\begin{figure}
\centering \epsfig{file=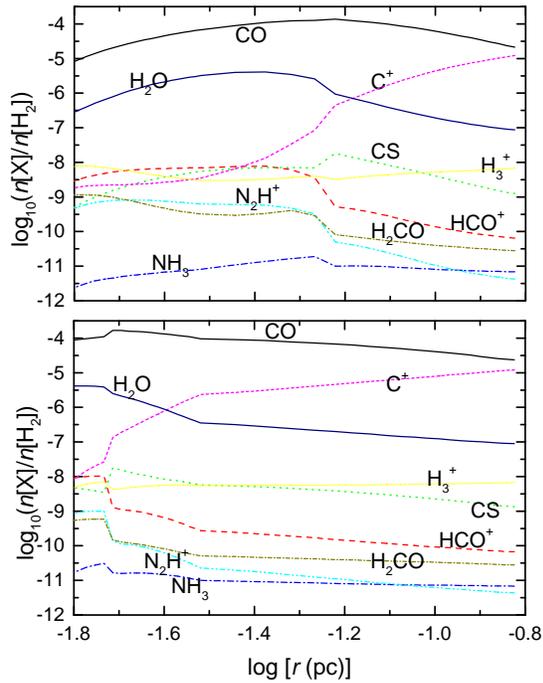, width=8 cm, scale=, clip=, angle=0}
\caption{The log fractional abundances of various species as a function of the
core radius at $t_{\rm coll}$ $=$ 1.36$\times$10$^5$\,yr ({\it top}) and
4.08$\times$10$^5$\,yr ({\it bottom}).}
\end{figure}

Having calculated the spatial distribution of the chemical abundances, we apply
an appropriate radiative transfer model to be able to predict the observed line
profiles. In this study we use {\sc smmol}; a one-dimensional ALI code that
solves multi-level non-LTE radiative transfer problems and which was described
in RY01. Subsequent to that study, {\sc smmol} has been benchmarked for
accuracy (van Zadelhoff et al. 2002) and expanded to include a more
comprehensive dataset of molecular collisional data. Defining an input
continuum radiation field, the code calculates the total radiation field and
the level populations, $n_{i}$, using a user-defined convergence criterion
which we set to $\Delta n_{i}/n_{i}\leq 10^{-4}$. The emergent intensity
distributions are then convolved with the telescope beam, so that the model
directly predicts the spectral line profiles for a given source as observed
with a given telescope. For the background radiation field we use the cosmic
background continuum ($T_{\rm BB}$ = 2.72\,K). We assume that the telescope
beam can be approximated by a Gaussian, with a characteristic half power beam
width (HPBW). In this paper we have concentrated on line profiles for the five
lowest transitions of \hcop\ and CS ($J$ $=$ 1$\to$0 to $J$ $=$ 5$\to$4), and
the \nthp\ $J$ $=$ 1$\to$0 line (its hyperfine structure was, however, not
considered; see Section~3.5). Data for \hcop\ and \nthp\ -- H$_2$ collisions
were taken from Flower (1999) and for CS -- H$_2$ collisions from Turner et al.
(1992).

%; and for o-\amm\ -- p-H$_2$ collisions from Danby et al. (1988).}

For a typical low mass star-forming region, the angular resolution of a single
dish mm/sub-mm telescope is comparable to the angular size of the core. For our
calculations we have adopted the physical parameters of the B335 core, as
deduced from Zhou et al. (1993), but placing it at 500\,pc (in the outskirts of
the Gould Belt or at roughly the distance of the Orion molecular cloud). For
the telescope beam we have used characteristics of a 45-m antenna corresponding
to the Nobeyama class. We have used beam efficiency data from commonly used
systems based on the Nobeyama 45-m, the IRAM 30-m, and the James Clerk Maxwell
Telescope (JCMT) 15-m antennae and modelled these as a putative 45-m class
telescope (putative in the sense that no 45-m dish exists that can fully sample
the frequency range that covers all of the transitions that we have
considered). The efficiencies across the frequency range thus correspond to
Nobeyama ($<$145 GHz), IRAM (145 -- 280 GHz) and the JCMT ($>$280 GHz). Beam
efficiencies at line frequencies are interpolated values generated by the {\sc
smmol} code. For the range of frequencies that we considered the spatial
coverage of the source is rather similar to that obtained with the JCMT 15-m
antenna for a source distance of 250\,pc (cf. the study by RY01). In Section
3.4 we comment on the sensitivity of the simulated line profiles to the adopted
distance of the cloud. The beam was placed at the core's centre in all cases
and no spatial offsets were considered. The telescope/beam characteristics and
utilized transitions are summarized in Table~2.

\begin{figure*}
\centering \epsfig{file=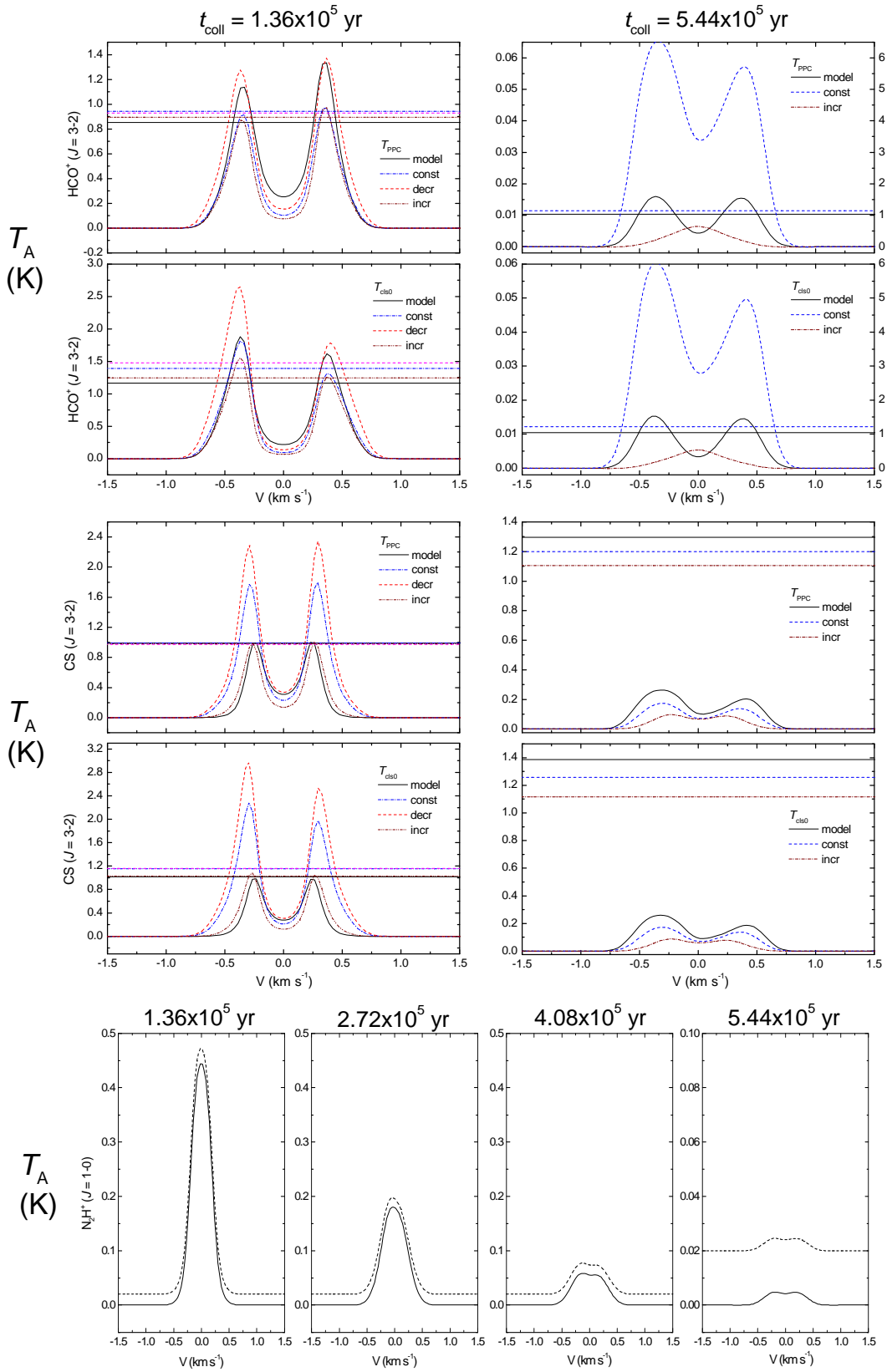, width=16 cm, scale=, clip=, angle=0}
\caption{\hcop\ ($J$ $=$ 3$\rightarrow$2) and CS ($J$ $=$ 3$\rightarrow$2)
spectra for the \tp\ and \tc\ temperature profiles, at early and late collapse
times \tcol\ $=$ 1.36$\times$10$^5$ and 5.44$\times$10$^5$\,yr, respectively.
The various profiles correspond to the `model', `constant', `decreasing', and
`increasing' abundance distributions discussed in the text; the horizontal
lines denote the respective blue/red asymmetry ratio (see Table A1; read off
the right-hand axis for the two top-right \hcop\ panels). The bottom four
panels show the evolution of simulated \nthp\ ($J$ $=$ 1$\rightarrow$0)
profiles for the \tp\ and \tc\ profiles across all collapse times adopting
`model' abundances divided by a factor of nine (solid and dashed line
respectively; for clarity the latter was scaled up in intensity by $+$0.02\,K.
See Section 3.5 for details).}
\end{figure*}

\begin{figure*}
\centering \epsfig{file=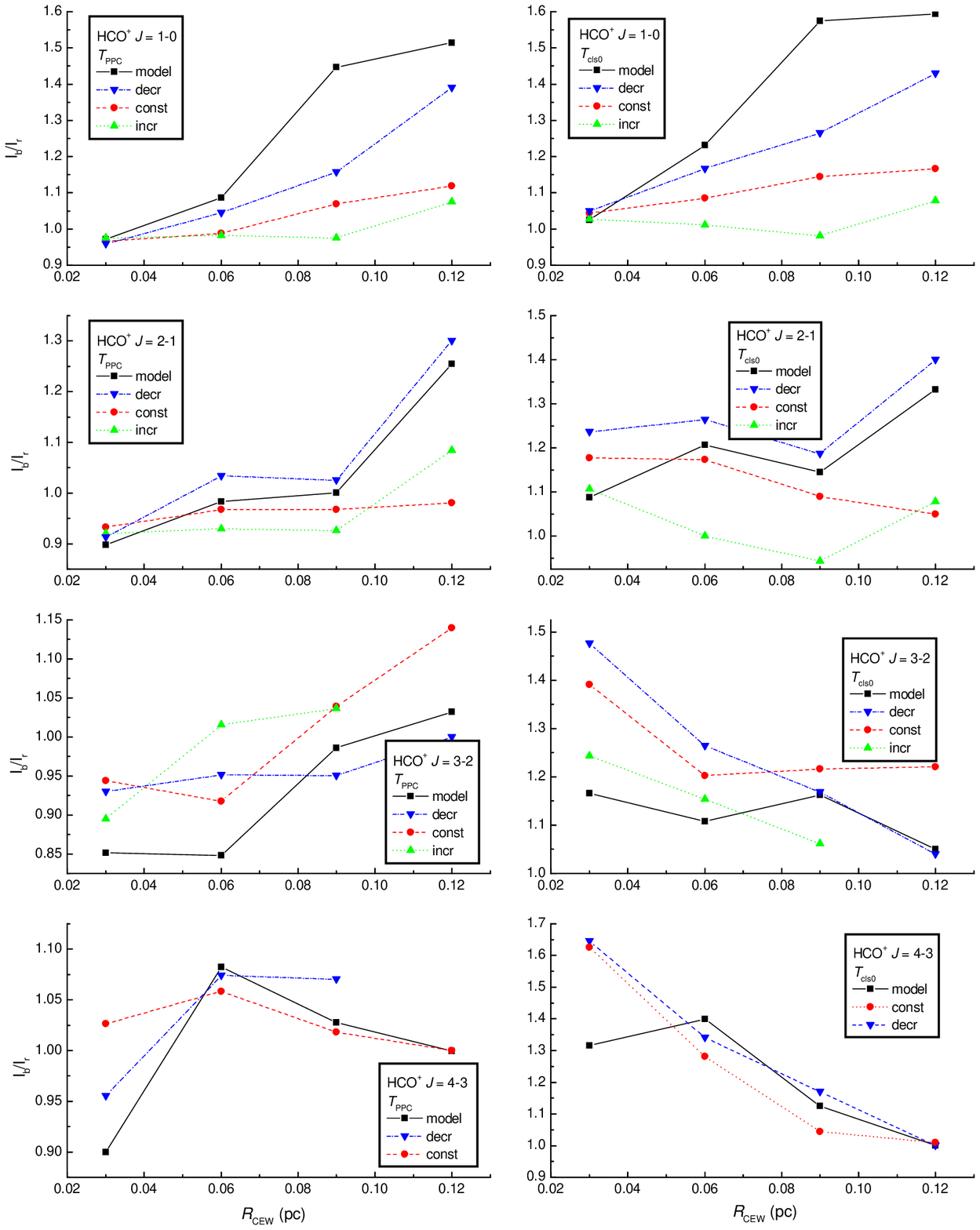, width=17.2 cm, scale=, clip=, angle=0}
\caption{Evolution of the blue- to red-shifted line asymmetry for \hcop\ ($J$
$=$ 1$\rightarrow$0, 2$\rightarrow$1, 3$\rightarrow$2, 4$\rightarrow$3)
transitions for models with \tp\ and \tc\ temperature distributions.}
\end{figure*}

\section{Results and discussion}

Raw results from the chemical models are given both in the form of a plot of
the {\it linear} abundances as functions of position at a single time
(Fig.\,~2) -- which emphasizes the relative contributions of the different
parts of the core to the line profiles -- and a more conventional plot of the
(logarithmic) abundances as functions of position at individual collapse times
(Fig.\,~3). These represent the outputs obtained from the (independently
integrated) Lagrangian grid of points.

Examples of the spectral line profiles for HCO$^+$, CS and \nthp\ calculated
using these abundances, the physical parameters obtained from the dynamical
model and appropriate source/telescope parameters are shown in Fig.\,~4 (and
also Fig.\,~12 -- see below). For each of these line profiles the continuum
(deriving from a combination of dust emission and the cosmic background
radiation) has been subtracted so as to allow an easier comparison of the
profiles. Line profiles are displayed for the first and last collapse periods
for the \hcop\ and CS $J$ = 3$\to$2 transitions.

The full set of results is given in numerical form in Table~A1 of the Appendix.
In constructing this table, and for the sake of brevity, we have reduced
individual line profiles to the peak line intensities of the blue-shifted
components ($I_b$) and the ratio of blue- to red-shifted component intensities
($I_b/I_r$). The tabulated intensities were determined by fitting Gaussians to
the simulated line profiles as if they were observed spectra. We refer to the
quantity $I_b/I_r$ as the line asymmetry ratio (or blue excess). This ratio is
also marked by the horizontal lines in Fig.\,~4. Inspection of this figure
allows the evaluation of the dependence of the line strength and asymmetry on
the adopted abundance distributions in the core.

For almost all transitions and collapse periods considered in this study the
peak line intensities, as well as the integrated fluxes, are the weakest for
abundance distributions that increase outwards from the centre of the core.
This can be explained in terms of a smaller optical depth and hence weaker
emergent line intensity for regions of the core where the mass-averaged
abundance of a molecular tracer is particularly low. For example, in the case
of \hcop\ at \tcol\ $=$ 1.36$\times$10$^5$\,yr this is true between $\sim$ 0.03
-- 0.09\,pc, and the line intensities corresponding to the `increasing'
abundance profile are typically weaker than for the other profiles. For CS the
`model' and `increasing' abundance profiles show similar, radially increasing
mass-averaged distributions out to $\sim$ 0.12\,pc. The line intensities for
these profiles are lower than for the constant and `decreasing' profiles which
are associated with larger mass-averaged abundances over the same region. Also,
the difference in line intensities from the adoption of the `increasing' as
opposed to, for example, the `decreasing' abundance profile is more pronounced
for lines of higher critical density (\ncr) and smaller HPBW as these are less
efficiently excited at large core radii where the gas density is low.

In Figs.\,~5 and 7 we present plots of the line asymmetry ratio (using some of
the values given in Table~A1) for observationally interesting transitions of CS
and \hcop\ respectively, showing the temporal evolution of the line asymmetry
ratio for models with a \tp\ and \tc\ temperature distribution. We can deduce
several trends from these results and discuss them in the following
subsections.

\subsection{\hcop}

\begin{figure}
\centering \epsfig{file=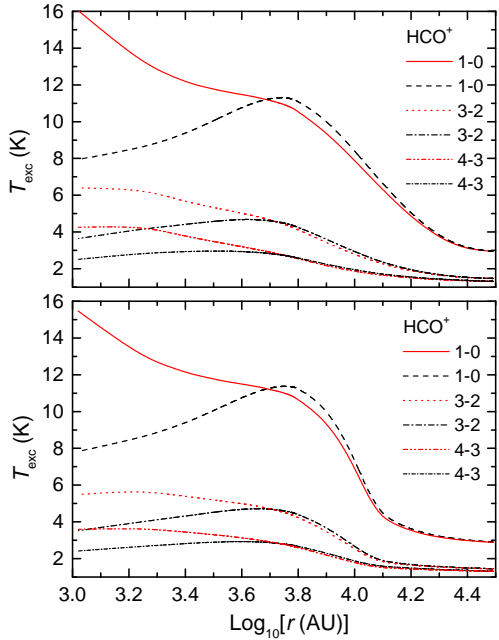, width=7.5 cm, scale=, clip=, angle=0}
\caption{Excitation temperatures of \hcop\ transitions corresponding to $t_{\rm
coll}$ $=$ 1.36$\times$10$^5$\,yr: (Top) adopting the `decreasing' abundance
distribution; (Bottom) adopting the chemical model abundance distribution. Red
curves are from the radiative transfer solution applicable to the \tc\
temperature profile and black curves are from the solution applicable to the
\tp\ profile.}
\end{figure}

In our study we find that for both the \tp\ and \tc\ temperature distributions
the line asymmetry ratio generally increases with time for the $J$ = 1--0
transition for the `model' and `decreasing' abundance distributions; it shows a
very slight increase for a constant molecular abundance, while it remains
mostly flat with the `increasing' abundance distribution. In the latter case
the line is almost symmetric, but the line does show a slight red asymmetry
($I_{\rm b}$/$I_{\rm r}$ $<$ 1) in models with a \tp\ temperature distribution:
the symmetric profiles are due to the decreased line opacity resulting from the
depressed molecular abundance towards the core's centre. The slight red
asymmetry is caused by a positive outward gradient of the line excitation
temperature (\tex) given the \tp\ profile in the inner core. The blue asymmetry
shows a rather stronger sensitivity on the abundance distribution for the 1--0
than for the 2--1 transition, especially at later times (Fig.\,~5).

The 2--1 line overall shows similar behaviours to that of the 1--0 line with
rising asymmetry ratios at later times (except for a constant and `increasing'
abundance in the \tc\ case): the asymmetry ratios however are generally
\emph{lower} and this is caused by the smaller HPBW of this transition which
covers about half the collapsing area of the cloud during each collapse period
(relative to the 1--0 line), and the higher critical density of this line
compared to the 1--0 transition (1.1$\times$10$^{6}$ vs.
1.6$\times$10$^{5}$\,\cmt). For the \tc\ temperature profile the asymmetry
ratios of the 2--1, 3--2 and 4--3 lines initially (at 1.36 $\times$ 10$^5$ yr)
attain higher values than that of the 1--0 line. This is due to the smaller
HPBWs of these lines which during that time respectively cover the central 38,
25, and 19 per cent of the infalling region thus sampling the warmest, dense
inner regions of the cloud.

In Fig.\,~6 we show an example of the radiative transfer solution for the
excitation temperature of representative \hcop\ lines for both our adopted
temperature profiles and for two different abundance distributions. For a given
temperature profile the \tex\ variations per line across the cloud appear
similar, but the temperatures are all slightly higher in the top compared to
the bottom panel along points in the cloud where the `decreasing' abundance
distribution results in a larger concentration of \hcop\ than the `model'
abundance distribution. The opposing trends caused by the two different kinetic
temperature profiles (Fig.\,~1) are especially evident towards the centre of
the cloud. When adopting the \tp\ profile the \tex\ of the lines shows a
positive outward gradient in the inner cloud which results in red asymmetry
ratios in the early collapse stages (Fig.\,~5). For the same temperature
profile, the lines attain blue asymmetry ratios during the late collapse stages
as the expanding CEW encounters a negative \tex\ outward gradient (for log($r$)
$\gtrsim$ 3.8): the 1--0 and 2--1 lines then reach slightly higher $I_{\rm
b}$/$I_{\rm r}$ ratios than the 3--2 and 4--3 lines and this must be caused by
the better sampling (larger HPBW) of the infalling region, combined with an
increased line opacity for the former diagnostics (especially when the `model'
and `decreasing' abundance distributions are adopted).

For the \tc\ temperature profile, the 3--2 and 4--3 lines initially show blue
asymmetries ($I_{\rm b}$/$I_{\rm r}$ $>$ 1) with a spread depending on the
adopted abundance distribution. This affects differently the line opacity in
each case, but as the collapse progresses the line profiles become more
symmetric: due to the central placement of the beams these high critical
density lines (3.4$\times$10$^6$ and 9.1$\times$10$^6$\,\cmt respectively)
cover at each successive collapse period a decreasing fraction of the infalling
gas volume whose density steadily falls. At \tcol\ $=$ 5.44 $\times$ 10$^{5}$
yr only 6 and 5 per cent of the infalling region is respectively sampled by the
3--2 and 4--3 lines compared to 19 per cent for the 1--0 line, and the central
density has fallen below 5$\times$10$^4$\,\cmt.

The $J$ $=$ 5--4 line is very weak for both adopted temperature profiles (due
to the high critical density for the transition): it shows no central
self-absorption in the \tp\ case, while it shows strong blue asymmetry at early
times in the \tc\ case, becoming symmetric later on (Table~A1).

\subsection{CS}

\begin{figure*}
\centering \epsfig{file=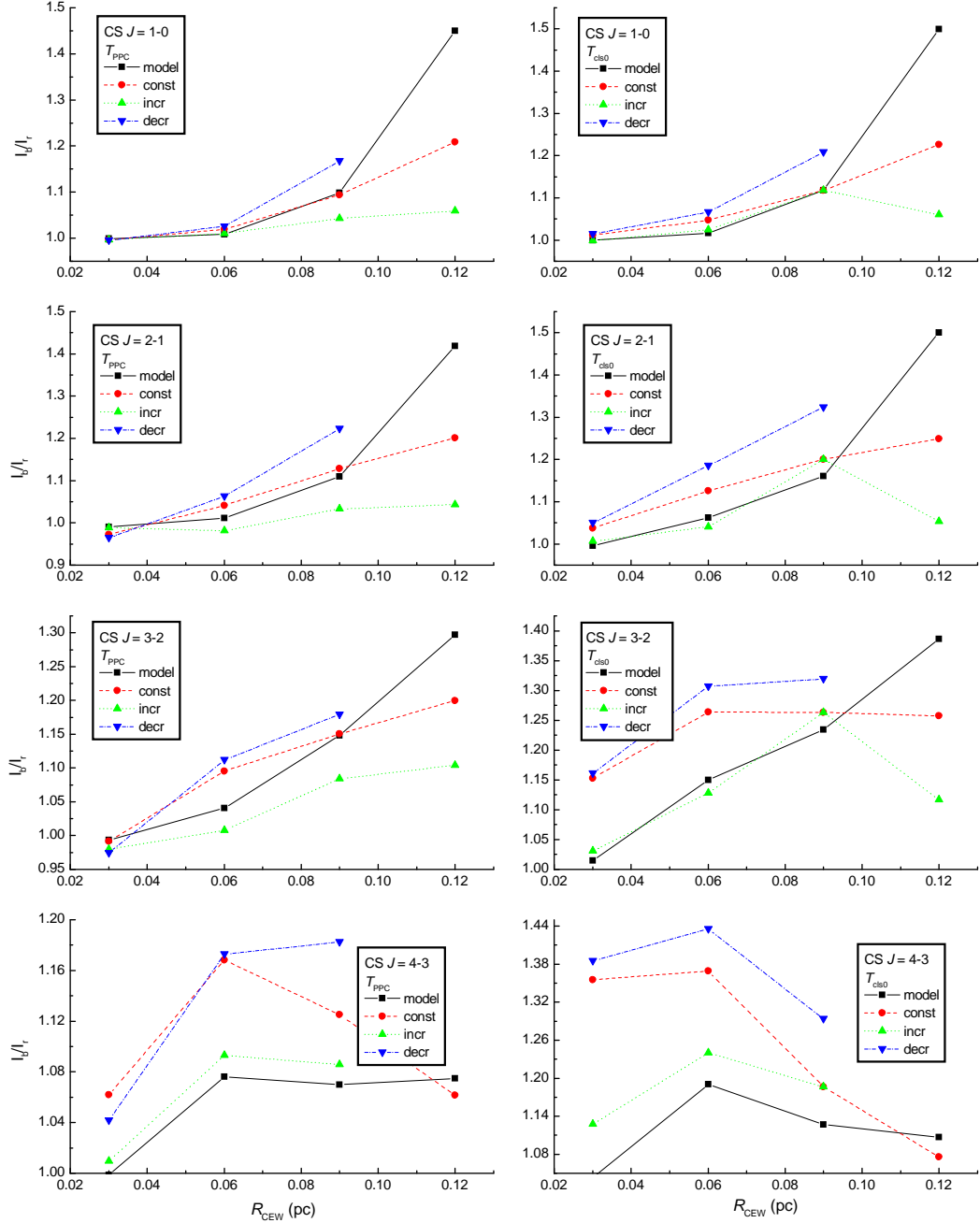, width=17.2 cm, scale=, clip=, angle=0}
\caption{Evolution of the blue- to red-shifted line asymmetry for CS ($J$ $=$
1$\rightarrow$0, 2$\rightarrow$1, 3$\rightarrow$2, 4$\rightarrow$3) transitions
for models with \tp\ and \tc\ temperature distributions.}
\end{figure*}

The $J$ $=$ 1--0, 2--1 and 3--2 transitions consistently show blue asymmetric
profiles whose asymmetry increases at later times with both \tp\ and \tc\
temperature distributions, without a strong sensitivity on the adopted
abundance for the species (Fig.\,~7). The only line that shows an opposite
trend is the 4--3 transition which shows a blue asymmetry at \emph{early} times
(most clearly in the models with a \tc\ temperature profile), which becomes
less prominent at later times. It is also strongly dependent on the adopted
abundance distribution. This is also true for the 5--4 transition; this is
always a very weak line. Compared to \hcop, the difference between the line
asymmetry resulting from the adoption of different kinetic temperature profiles
is not so pronounced, although for the \tc\ models slightly higher blue
asymmetries are obtained. How can the different behaviour of CS compared to
\hcop\ be explained? A plot of \tex\ for CS transitions reveals a broadly
similar picture to Fig.\,~6 and the contrasting trends corresponding to the
\tp\ and \tc\ cases remain. The outward gradient in the \tex\ of all analyzed
lines when adopting a \tp\ profile is however only slightly positive in
contrast to the case of \hcop.\footnote{For example, the excitation temperature
differential in the \tp\ case for log($r$) $=$ 3 -- 3.8 is $<$ 2\,K for the CS
1--0 line compared to $\approx$ 4 K for \hcop\ 1--0.} Furthermore, all the CS
lines have larger HPBWs than the corresponding \hcop\ transitions with our 45-m
antenna set-up. As a consequence, when adopting the \tp\ profile, the
transitions plotted in Fig.\,~7 are all initially (at 1.36 $\times$ 10$^5$ yr)
approximately symmetric, in contrast to the \hcop\ case where slightly negative
asymmetry ratios are obtained. At later times as the beam footprint starts
sampling infalling regions where the \tex\ of the lines through the cloud
becomes centrally peaked, the $I_{\rm b}$/$I_{\rm r}$ ratios typically
increase. The high \ncr\ lines 4--3 and 5--4, which incidentally also have the
smallest HPBWs, do not follow this trend at late collapse times because by then
the core density is low and the lines are not sufficiently opaque compared to
the lower $J$ transitions.

The 5--4 line becomes very weak after the first collapse period (Table A1) as
it has a high critical density (9 $\times$ 10$^6$~\cmt) and cannot be
efficiently excited in the decreasing density of the infalling envelope at
later times. As CS is depleted in the chemical model in the inner core, the
line appears optically thick and self-absorbed only for the constant and
decreasing abundance distributions with a blue asymmetry that decreases rapidly
after the first collapse age (as in \hcop\ 4--3); again \tc\ models show higher
blue asymmetry ratios than \tp\ models.

\subsection{Chronometers of inside-out collapse?}

\begin{figure}
\centering \epsfig{file=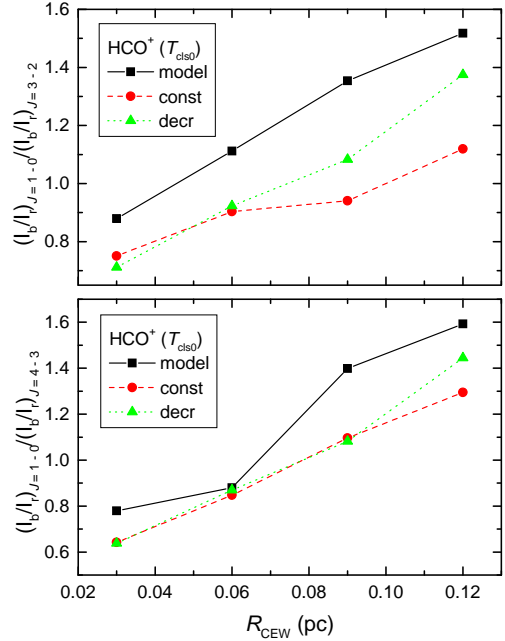, width=7.5 cm, scale=, clip=, angle=0}
\caption{Evolution of the {\it relative} blue excess (the ratio of the
respective line asymmetries) as function of the outer collapse expansion wave
radius for \hcop, in a \tc\ model, for the `model', `constant', and
`decreasing' abundance distributions in the core: (Top) $J$ $=$ 1$\rightarrow$0
{\it vs.} 3$\rightarrow$2; (Bottom): $J$ $=$ 1$\rightarrow$0 {\it vs.}
4$\rightarrow$3 transitions. See also Table~A1.}
\end{figure}

\begin{figure}
\centering \epsfig{file=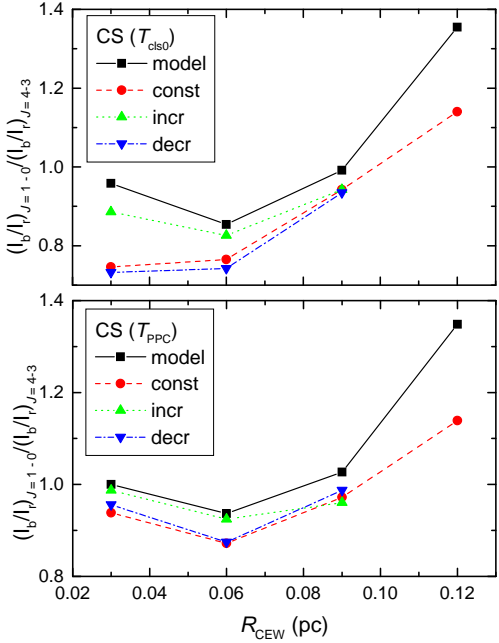, width=7.5 cm, scale=, clip=, angle=0}
\caption{As in Fig.\,~8 but for CS $J$ $=$ 1$\rightarrow$0 {\it vs.}
4$\rightarrow$3 transitions: (Top) model core with a \tc\ temperature profile;
(Bottom) model core with a \tp\ profile.}
\end{figure}

\begin{figure}
\centering \epsfig{file=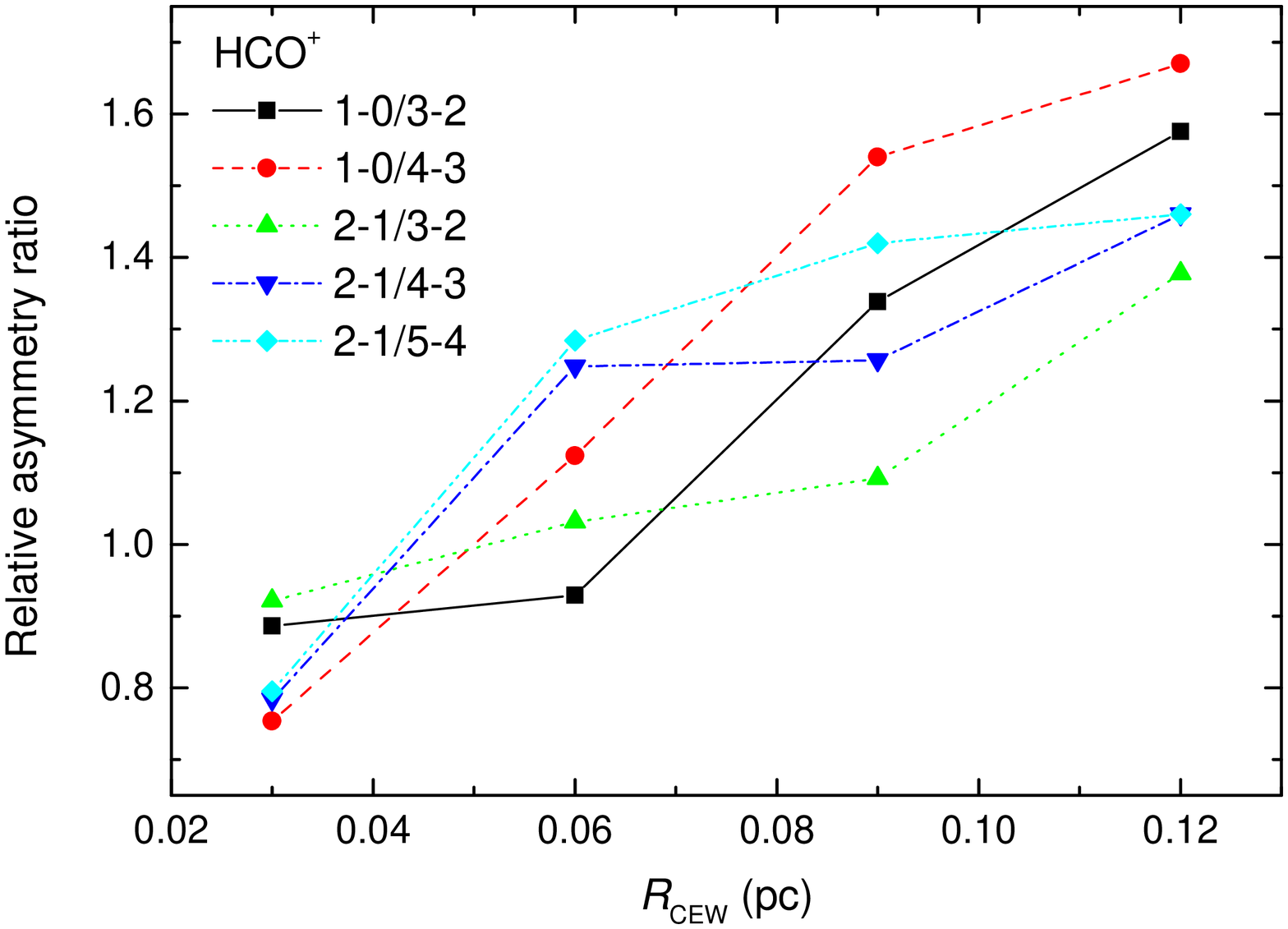, width=6.6 cm, scale=, clip=, angle=-90}
\caption{As in Fig.\,~8 but for the relative asymmetry ratios of the \hcop\
lines noted in the key. The simulation is for a 15-m antenna for a core with a
fixed \tc\ temperature profile and a `model' \hcop\ abundance distribution at a
distance of 250\,pc.}
\end{figure}

\begin{figure}
\centering \epsfig{file=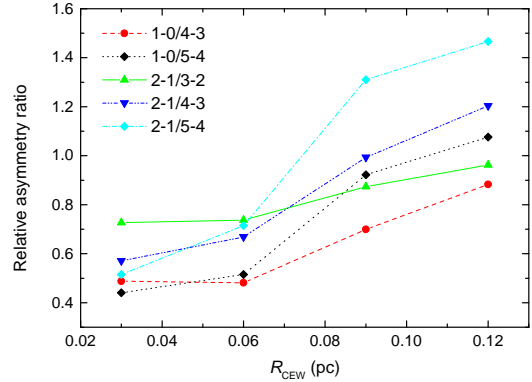, width=6.6 cm, scale=, clip=, angle=-90}
\caption{As in Fig.\,~8 but for the relative asymmetry ratios of the \hcop\
lines noted in the key. The simulation is for a 45-m antenna for a core with a
fixed \tc\ temperature profile and a constant \hcop\ abundance distribution at
a distance of 500\,pc.}
\end{figure}

Studying the above results, we see that the \hcop\ $J$ $=$ 3--2, 4--3, and 5--4
transitions follow an inverse trend in the temporal evolution of their blue
asymmetry compared to the 1--0 transition. To help visualize these effects we
plot in Fig.\,~8 the time evolution of the \emph{relative} blue excess;
\[\mathcal E =
\frac{(I_b/I_r)_{1\to0}}{(I_b/I_r)_{4\to3}},~\frac{(I_b/I_r)_{1\to
0}}{(I_b/I_r)_{3\to2}}. \] Results are shown for models with a \tc\ temperature
profile where effects are most noticeable. Similarly, in Fig.\,~9 we plot the
evolution of the relative blue excess, $\mathcal E$, of the CS $J$ $=$ 1--0
{\it vs.} the 4--3 transition (for models with \tp\ and \tc\ distributions). It
can be seen that the $\mathcal E$ corresponding to these transitions depends on
the dynamical age, and models with different abundance distributions follow the
same general trend. A similar dependence is the one emerging from a comparison
of the contrasting evolution of the \hcop\ 4--3 vs. the CS 3--2 line (compare
Figs.\,~5 and 7, for a core with a \tc\ temperature profile and a `model'
abundance distribution).

These behaviours are mostly due to the varying spatial sampling of parts of the
core at various infall times by the different line frequencies employed, given
the adopted 45-m antenna and the centrally placed beams. For instance, the
\hcop\ $J$ $=$ 1--0 line samples 76 per cent of the infall region at $t_{\rm
coll}$ $=$ 2.72$\times$10$^5$\,yr, when the 3--2 and 4--3 transitions sample 25
and 19 per cent of its angular size respectively (using the parameters in
Tables~1 and 2). In trials where the 3--2 and 4--3 transitions were modelled
with a beam size that matched that of the 1--0 line, i.e. when employing a
11--15-m size antenna for the former lines and a 45-m antenna for the latter
line, the evolution of the asymmetry for all these lines was roughly similar,
and the trends in Fig.~\,8 virtually disappeared. The same case can be made for
the CS trends of Fig.\,~9.

In order to test whether these results are the product of a particular
modelling configuration, in Fig.~\,10 we plot the relative line asymmetry,
$\mathcal E$, of \hcop\ transitions modelled for a 15-m antenna (JCMT
equivalent) and for a source distance of 250\,pc. This results in approximately
the same spatial coverage of the core by the various transitions as when the
standard values of Table~2 are used. The `model' \hcop\ abundance distribution
and the \tc\ temperature profile were adopted. The dependencies on the collapse
period noted above hold in this case too. This shows that as long as the
spatial coverage of the core remains approximately the same, the choice of
modelling configuration does not significantly affect the trends of Fig.\,~8.

In order to further check whether these trends might still be valid for a much
denser core model, we performed a simulation arbitrarily increasing the density
across the core twenty-fold, up to a peak density of 3.6$\times$10$^6$\,\cmt\
at the onset of infall. This value is near the top of the range of densities
inferred for a sample of protostellar cores in the Lynds clouds (Visser et al.
2002) and in Perseus (Motte \& Andr\'e 2001); although obviously unphysical in
the context of the inside-out collapse model, this simulation serves to
highlight the sensitivity to density. The relative line asymmetry ratios for a
set of \hcop\ transitions were obtained after evolving the cloud through the
same collapse periods as for our standard 45-m antenna models (control
parameters of Table 2). The results are plotted in Fig.\,~11 and show that the
dependencies identified above are still valid.

%In contrast to the behaviour of our standard models of Fig.\,~5, the asymmetry
%ratio of the 1--0 line now barely increases, whereas that of the 2--1 line
%\emph{decreases} during the same period (only {\em relative} line asymmetries
%are shown -- not the nominal blue/red asymmetry ratios). At the same time the
%high critical density 4--3 and 5--4 lines become stronger \emph{and} show a
%steeper decline of their asymmetry ratio starting from higher values than
%before. As a result, for the full duration of the collapse, the relative line
%asymmetry ratio $\mathcal E$(2$\to$1/4$\to$3) for example, now encompasses the
%same range as the $\mathcal E$(1$\to$0/4$\to$3) ratio in the standard model.}

Clearly, the trends in Figs.\,~8--11 imply that the behaviour of the
\emph{relative} blue excess of certain molecular transitions (observed with a
given antenna) can be considered as a collapse chronometer in the formal
context of the Shu model. We have therefore shown that in idealized
spherically-symmetric cores which possess inward only motions, line
measurements coupled to detailed models would in theory be able to constrain
the evolutionary state of the (assumed inside-out) collapse. A number of
caveats however must accompany this statement.

Firstly, the question of timescales. Observationally, the lifetime of
submillimetre-detected, pre-stellar cores has been estimated from the
statistics of detections of sources and approximate ages of 0.3\,Myr have been
established from an ensemble of various star-forming regions (Kirk et al.
2005). Based on estimates of the lifetime of Class~I sources, and using
statistical methods, a reference value for starless cores (PPCs) is $\sim$
0.6\,Myr (Kirk et al. 2007). In Perseus the lifetime of submm-detected starless
cores was also estimated to be $\sim$ 0.4 Myr (Hatchell et al. 2007).
%The timescales we considered in this work are approximately those.
The lifetime of Class~0 sources, which are potential manifestations of the
inside-out collapse paradigm, has been revisited: whereas it was previously
estimated that it may be one order of magnitude smaller than that of Class~I
sources, and of the order of 10$^4$ yr (e.g. Ward-Thompson 1996), it is now
thought that the relative Class~0/I phase lifetimes are similar (in the Perseus
and Lynds dark clouds at least, and assuming a constant star formation rate),
and of the order of a few $\times$ 10$^5$ yr (Visser et al. 2002; Hatchell et
al. 2007). The infall timescales we considered fall in this range. Moreover,
the boundary between the Class~0/I phases in terms of \emph{observed}
evolutionary indicators is still elusive (Hatchell et al. 2007). We thus deem
that, for the purposes of this work, the choice of timescales for our
collapsing core analogue is not without due justification.

Another caveat involves the temperature profiles we have adopted in our models
and which remain constant throughout the collapse phase. In realistic
situations one would expect some overall heating to ensue at later times as
mass accretes onto a central protostar. In order to get an idea of how this
change in temperature might affect the line asymmetry, we run a model adopting
a radial temperature profile, $T$($r$) $=$ 20($r$/1000 au)$^{0.36}$~K,
corresponding to a source of $\sim$ 15 L$_\odot$ (taken from Ward-Thompson \&
Buckley 2001). Our typical Class~0 temperature profile (Fig.\,~1) is from the
best-fit spectral energy distribution of B335 for a luminosity of $\sim$ 3
L$_\odot$ (Shirley et al. 2002) and so, in an approximate way, the above
$T$($r$) should represent a considerable `evolution' towards the Class~I
boundary. We find that for a \emph{constant} \hcop\ fractional abundance of
10$^{-8}$ the asymmetry ratio of \hcop\ $J$ $=$ 1--0 rises more steeply than
the one shown in Fig.\,~5 (top-right panel) corresponding to a model with a
\tc\ temperature profile. If we were to assume that heating takes place after
the second collapse period then the rise of the line asymmetry would be even
more pronounced. The blue excess of the 4--3 line decreases linearly with
collapse period, similarly to the model with a \tc\ profile, but attains
slightly higher values throughout. We therefore conclude that the trend shown
in Fig.\,~8 (bottom) would be little affected with this change of temperature
law. Moreover, the line asymmetry of the 5--4 line is negligibly affected, and
the 1--0/5--4 relative asymmetry ratio shows a dependence with dynamical age
similar to those of Fig.\,~8 as well.

Finally, a stronger caveat involves the possible generic applicability of the
identified `chronometers' in studies of realistic infall candidates. As noted
in Section~1, unfortunately, protostellar cores do not universally adhere to
the inside-out paradigm of collapse. Observed non-zero velocities at large
radii and/or small infall velocities at small radii, deviations from spherical
symmetry, outflows, and rotation, are all likely to affect the line asymmetry
of tracer species in ways that would be particular to each source. So would
observational uncertainties, such as beam pointing inaccuracies. However, the
identified trends are valid in the strict context of the inside-out collapse
model and give us valuable insight into the ways the line asymmetry evolves and
into the main factors that drive it.

In the next paragraph we examine the line profile sensitivity to the distance
from the source.

\subsection{Sensitivity to distance}

\begin{figure*}
\centering \epsfig{file=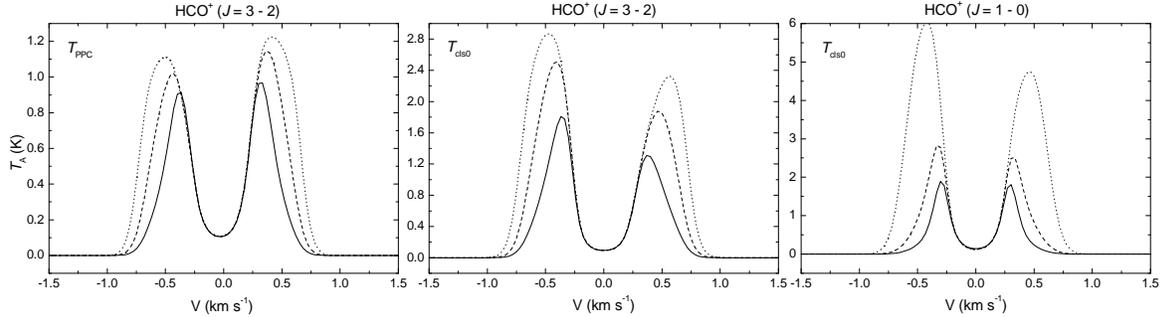, width=11 cm, scale=1., clip=, angle=-90}
\caption{Illustration of the dependence of the simulated line profile on the
assumed distance to the a source with a \tp\ or \tc\ temperature distribution.
Spectra from a 45-m antenna are plotted for distances of 50 pc (dotted lines),
250 pc (dashed lines) and 500 pc (solid
lines).} %bbllx=-100, bblly=-50, bburx=580, bbury=900
\end{figure*}

The sensitivity of the line asymmetry ratio to source distance is a complex
issue.
%We find that for a given antenna diametre the line
%asymmetry ratio is a stronger function of core distance for transitions which
%have a larger HPBW (towards lower frequencies).
For our nominal 45-m antenna the blue/red asymmetry of the \hcop\ $J$ $=$ 1--0
transition increases by 22 per cent when the distance is decreased from 500 to
50\,pc, whereas that of the $J$ $=$ 3--2 and 4--3 transitions decrease by
$\sim$ 12 and 29 per cent respectively (for models with a \tc\ temperature
profile). A possible explanation for these trends may be that the low \ncr\
1--0 line samples better the denser infall region of the core at smaller
distances than farther away, whereas the opposite is true for the higher \ncr\
3--2, 4--3 lines which at small distances sample a very small fraction of the
infall region (less than five per cent of the region within the 0.03 pc radius
of the collapse expansion wave). A similar variation in absolute values is seen
when adopting a 15-m antenna, but in that case the asymmetry ratio of the 3--2
and 4--3 transitions is larger at 250\,pc.
%No clear trends with either the line critical density or the HPBW are
%apparent.

An intrinsic property of the inside-out model seems to be that for whatever the
distance to a source of given density, a set of transitions can be found whose
optimum combination of HPBW (on a putative antenna), molecular tracer
abundance, and line \ncr, would allow them to act as collapse chronometers.

%For both antenna diametres the difference in the line asymmetry of these two
%transitions becomes the smallest at 50\,pc.
%the 1--0 line samples 152\% of this with the core at 500\,pc (with the 45-m
%dish), and 46\% of it with the core at 50\,pc (with the 15-m dish); for the
%3--2 line these values are 51 and 11\% respectively.

In Fig.\,~12 we plot the \hcop\ 1--0 and 3--2 spectra of a core for a number of
assumed distances, whilst keeping the beam size fixed (45-m antenna). They are
for a model with $R_{\rm CEW}$ $=$ 0.03\,pc and a constant molecular abundance.
The peak line temperatures for models with a \tc\ temperature distribution
decrease linearly with distance (apart from the 1--0 line whose peak
temperature shows a faster decrease). This is due to the fact that at near
distances the beam predominantly samples a denser portion of the core than at
larger distances where a bigger volume of gas is sampled. Also, at near
distances the peak centroids are found at higher velocities as the beam samples
a faster moving gas flow at the core's centre; a similar result was reported by
Ward-Thompson and Buckley (2001). The same comments apply in the case of a core
with a \tp\ temperature distribution.

We conclude that distance uncertainties for a given core are likely to affect
significantly the simulated line asymmetry ratios as these depend critically on
the coupling between the HPBW of the transition probe and the size of the
infall region.

%On the other hand, as the regions that give rise to the blue and red peaks of
%the line profile have the same temperature difference irrespective of the
%source's distance, the asymmetry ratio remains rather insensitive to the
%adopted distance.

\subsection{\nthp}

\begin{figure*}
\centering \epsfig{file=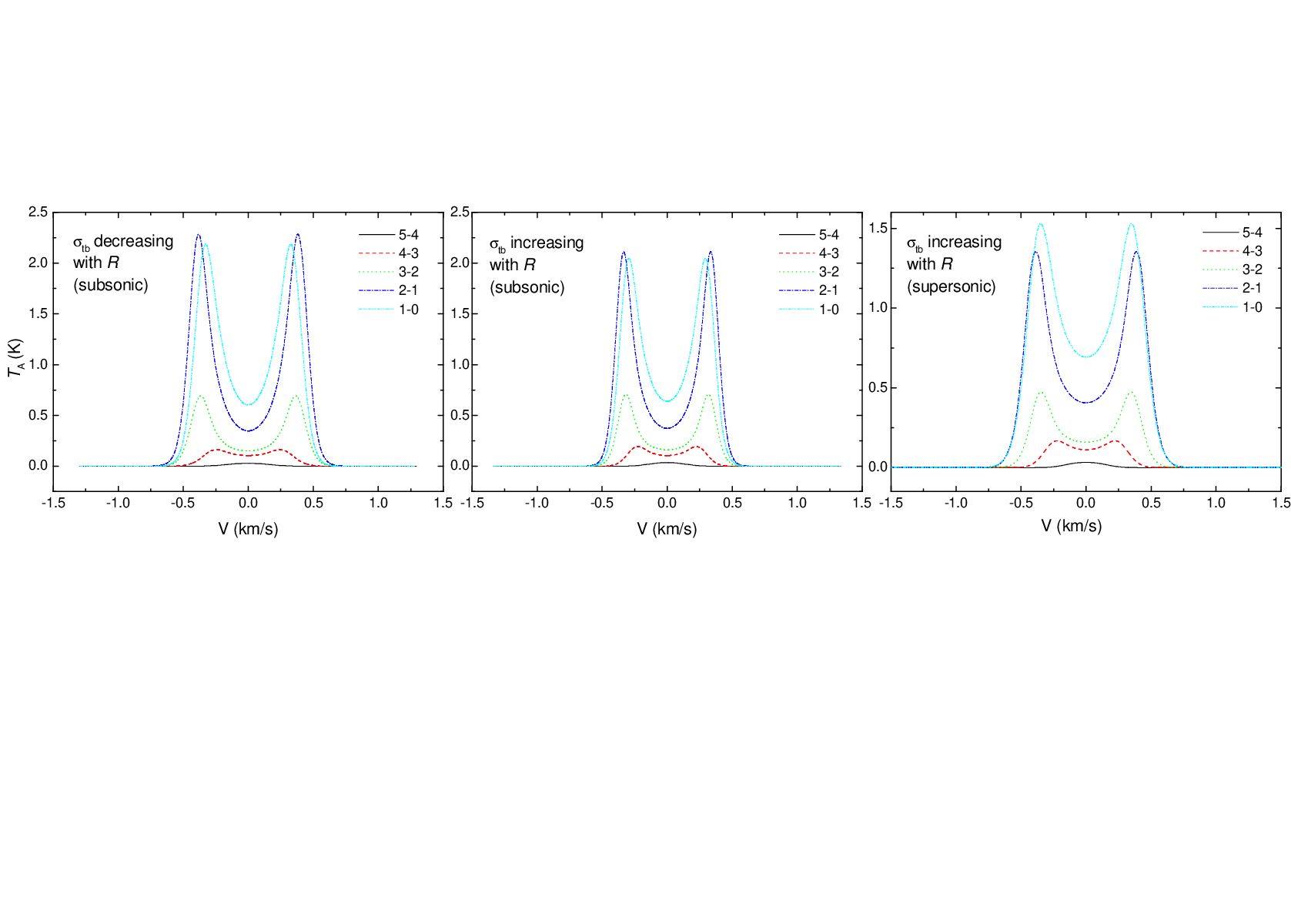, width=11 cm, scale=, clip=, angle=-90}
\caption{Illustration of the dependence of the simulated \hcop\ line spectra of
a static core on the assumed level of internal turbulence -- the \tc\
temperature profile was adopted with a molecular abundance distribution as that
shown in Fig.\,~2 (`model'); see text for details.}
\end{figure*}

The $J$ $=$ 1--0 transition of diazenilium has been modelled using only the
molecular abundance from the chemical code as input to the RT code. The mean
fractional abundance of \nthp\ in the chemical model displays a range of $\sim$
1$\times$10$^{-10}$ to 9$\times$10$^{-10}$ in the timescales we considered;
this is in very good agreement to the range found for a large sample of PPCs in
the Perseus molecular cloud (Kirk, Johnstone, \& Tafalla 2007). We have not
attempted to model the \nthp\ hyperfine structure (see Pagani et al. 2007). In
order to make our results more readily comparable to observations we have thus
divided the molecular abundance corresponding to each collapse period by a
factor of nine; this would approximately reproduce the flux of the isolated
`blue' hyperfine component $FF'$ $=$ 01--12 (assuming that the hyperfine levels
are populated in proportion to their statistical weights). This is usually
better suited for analysis of infall candidates as it is mostly optically thin
(Caselli et al. 1995; Mardones et al. 1997), and the other components can
suffer from overlap which occurs due to the large intrinsic linewidths in dense
cores resulting from non-thermal motions and opacity broadening. The line is
extensively used in studies of star-forming cores as the tracer whose largely
single-peaked nature argues against interpretations other than infall when
trying to explain the blue asymmetry exhibited by other tracers.

A graph of the evolution of the transition during the collapse of the core
analogue is shown in Fig.\,~4 (bottom). The line appears single peaked and
optically thin for both the \tp\ and \tc\ temperature profiles, but becomes
double peaked during the last two collapse periods when its peak intensity has
dropped to about 10 and 2 per cent of its initial value. As the line is still
optically thin in the two right panels of Fig.\,~4 (bottom) the double peak is
not produced by self-absorption but instead it directly traces the infall
velocity structure of the core. A comparison of the model spectra with the PPC
observations presented by Lee et al. (1999) shows very good agreement in terms
of peak intensity and line width with the range
%(0.05--0.40\,K and 0.20--0.55 km\,s$^{-1}$ respectively)
detected in their sample for the isolated component. At $t_{\rm coll}$ $=$
1.36$\times$10$^5$ yr the line characteristics resemble those of L183B, while
at 4.08$\times$10$^5$ yr the line looks very much like the one present in L492
(cf. fig.\,~5 in Lee et al.).
%; the latter source is also listed as a strong infall candidate by Sohn
%et al. (2007) with a gas velocity that probably increases towards the centre.
In the context of inside-out collapse, the suitability of this line in the role
of the optically thin tracer of infall (except perhaps in advanced stages) is
re-affirmed by this work.

\subsection{Microturbulence}

In this paragraph we comment on the effects of varying levels of internal
turbulence on the simulated spectra of a \emph{static} core with a \tc\
temperature profile. By setting the infall velocity to zero in these models the
effects of varying the turbulence are more readily observed. Ward-Thompson,
Hartmann and Nutter (2005) have shown how increased levels of \emph{constant}
turbulent velocity dispersion, $\sigma_{\rm tb}$, throughout an infalling
source lead to increased velocity separation of the line peaks. They reported
peak velocity differences (i.e. separation of the blue/red peaks) of order
$\Delta V$ $\sim$ 1--2 km s$^{-1}$ when the turbulent velocity varied between
1/3.5 and 1.14 times the adopted sound speed, $\alpha_{\rm eff}$, in a
collapsing cloud.

We have investigated plausible \emph{monotonic variations} of $\sigma_{\rm
tb}$, increasing or decreasing with distance from the static core's centre. We
found that when $\sigma_{\rm tb}$ decreases outwards from subsonic values of
1.5$\sigma_{\rm NT}$ to $\sigma_{\rm NT}$ at the outer core radius (where
$\sigma_{\rm NT}$ $\approx$ 0.145 km\,s$^{-1}$ is our nominal turbulent
velocity dispersion), this produces a peak velocity separation of 0.7 and 0.8
km s$^{-1}$ for the \hcop\ $J$ $=$ 1--0 and 2--1 transitions respectively
(Fig.\,~13). When $\sigma_{\rm tb}$ instead increased with radius by the same
amount (remaining subsonic), the peak velocity separation of the $J$ $=$ 1--0,
2--1, and 3--2 lines was $\sim$ 15 per cent smaller, while that of the 4--3
line changed by less than 10 per cent. Higher $J$ lines were not self-absorbed
and the full width at half maximum of the 5--4 line was 20 per cent larger in
the `decreasing' compared to the `increasing' $\sigma_{\rm tb}$ case (for the
same radiant line flux). The differences in the velocity separation of the line
peaks that we found in these two simple examples are similar in magnitude to
what one obtains if the optical depth of the lines are increased by a factor of
$\sim$2 as Ward-Thompson, Hartmann and Nutter report. However, this may be the
case only for small $\sigma_{\rm tb}$ variations, such as these in our example;
Ward-Thompson and Buckley (2001) found that various power-law relations between
$\sigma_{\rm tb}$ and the gas density, encompassing a wider range than we have
considered, have a more substantial effect on the peak velocity separation for
the  \hcop\ $J$ $=$ 4--3 and CS $J$ $=$ 5--4 lines.

We also note that uncertainties in the distance to the source can mimic the
effects on the peak velocity separation produced by the range of $\sigma_{\rm
tb}$ we have explored. The distance uncertainties to protostellar cores,
however, do not usually exceed a factor of two and so the effects are probably
separable. If one includes an analysis of optically thin lines as well, then
the potential degeneracy between distance and turbulence effects can be
avoided.

Supersonic turbulence in the outer parts of pre-stellar cores has been
associated with outflows or with generic turbulence permeating from the parent
molecular clouds. We have thus investigated an outward increase of turbulence
to supersonic values at the edge of the core ($\sim$ 2$\alpha_{\rm eff}$),
encompassing a factor of three variation from the inner to the outer radius.
This is similar to that observed in the Class~0 source NGC\,1333-IRAS\,4A
(Belloche, Hennebelle, \& Andr\'{e} 2006). The velocity separation of the peaks
in this case (Fig.\,~13 right) is similar to what is obtained with a radially
decreasing subsonic $\sigma_{\rm tb}$. This situation however is clearly
distinguishable from the subsonic models as the total flux in the self-absorbed
lines ($J_{\rm up}$ $\leqslant$ 4) is lower.
%for instance, the flux in the 2--1 line is 10 and 30 per cent less than in the
%`increasing' and `decreasing' subsonic $\sigma_{\rm tb}$ cases respectively.
In the supersonic turbulence case, the optical depth of the lines in the centre
of the core \emph{relative to} the outer regions is larger as increased levels
of turbulence in the outer core result in lower optical depths (as the optical
depth is inversely proportional to the line width; see also Ward-Thompson and
Buckley 2001); hence the emission in the self-absorption trough which is more
weighted towards the inner core, is enhanced relative to that in the peaks.

\section{conclusions}

Using a coupled chemodynamical and RT model we have conducted an exploratory
study of the evolution of the spectral line characteristics (intensities and
asymmetries) of molecular species such as \hcop, CS, and \nthp, which are
frequently used as tracers of collapse in star-formation studies. We have
investigated the sensitivity of the line diagnostics to several chemical
abundance distributions and to the thermal structure of a low-mass core
analogue undergoing inside-out collapse.

\setcounter{table}{2}
\begin{table}
\caption{Comparison of the mean `blue-excess' $I_{\rm b}$/$I_{\rm r}$ for
models with a \tp\ and \tc\ temperature distribution (from {\it all} abundance
distributions and transitions) at early (`1': $t_{\rm coll}$ $=$
1.36$\times$10$^5$\,yr) and late (`2': $t_{\rm coll}$ $=$
5.44$\times$10$^5$\,yr) collapse periods.}
\begin{center}
\begin{tabular}{|lll|}
\hline \noalign{\vskip2pt}
\multicolumn{3}{c}{\hcop} \\
\noalign{\vskip2pt}
\multicolumn{1}{c}{~~~{\it \tp}}  &{\it \tc} &Ratio\\
\noalign{\vskip2pt}
1.~~~0.936$\pm$0.043     &1.334$\pm$0.527 &1.425$\pm$0.441   \\
2.~~~1.221$\pm$0.308  &1.159$\pm$0.184 & 0.949$\pm$0.411   \\
\noalign{\vskip2pt}
\multicolumn{3}{c}{CS} \\
\noalign{\vskip2pt}
1.~~~1.012$\pm$0.050     &1.149$\pm$0.051 &1.135$\pm$0.094   \\
2.~~~1.188$\pm$0.163     &1.225$\pm$0.190 &1.031$\pm$0.292   \\
\noalign{\vskip2pt} \hline
\end{tabular}
\end{center}
\end{table}

In Table~3 we compare the mean line asymmetry ratio of models with \tp\ and
\tc\ temperature profiles (respectively applicable to starless cores and cores
with a warm central region), for the five lowest transitions of \hcop\ and CS,
at early and late collapse times (the errors are 1$\sigma$ standard
deviations). This rather crude comparison smoothes out the effects of different
abundance distributions, but helps us to see that at early times \hcop\ is
generally a more sensitive probe of collapse than CS (at least in \tc\ models),
mainly because its higher excitation transitions resolve the infall region
better than the respective CS lines. At later times the situation is mildly
reversed (for \tc), mainly because CS 3--2 and 4--3 trace relatively better the
low density infalling envelope. For both species the mean asymmetry is stronger
in \tc\ models than in \tp\ models at the onset of infall, but this difference
almost disappears later on. Also, while for CS the line asymmetry increases on
average from early to late times, the \hcop\ asymmetry for the \tc\ models
decreases on average in the same length of time.

When different abundance distributions are considered then the picture becomes
more complicated. Figs.\,~4 and 7 show that if the molecular tracer's abundance
increases towards large core radii then the lines can appear symmetric even at
advanced stages of infall. This is mostly evident in models with a \tp\
temperature profile which are not conducive to large blue excesses in general
and for which red asymmetry ratios can be obtained. Conversely, a molecular
abundance which declines with radius can result in high values of blue excess
-- this is particularly evident in the behaviour of some lines at early times.
Finally, a consequence of the specific chemical model used in this study is
that the existence of high \hcop/CS abundance ratios in the inner core
contributes, at early collapse times, to a lower line asymmetry for CS than for
\hcop\ lines.

The main conclusions of this analysis of Shu-type infall are:

1. Overall, \hcop\ is a somewhat better tracer of core collapse than CS, but
this depends on the evolutionary state of the core, as there is a direct
correspondence between the critical density of a particular transition and its
ability to trace gas of given density and temperature (with a significant
excitation temperature and optical depth). \nthp\ $J$ $=$ 1--0 is well suited
as the optically thin `control' line in infall studies (except perhaps in
advanced evolutionary stages).

2. In the context of inside-out infall (in the absence of rotation) the
\emph{relative} blue asymmetry of certain transitions is a function of the
dynamical age (for a given telescope), and acts as a `collapse chronometer'.
This is mainly due to a combination of beam filling factors and line critical
densities: a line of high critical density whose beamwidth efficiently resolves
the infall region shows higher blue asymmetry at early collapse times than at
later times when the gas density becomes too low; the opposite is true for a
line of low critical density whose large beamwidth cannot adequately resolve
the infall region at early times, but does so later on whilst remaining
sufficiently opaque in the lower density conditions.

3. The line asymmetry ratio is dependent on the abundance distribution of the
molecular tracer in the source. The dependence naturally becomes stronger when
the line beamwidth resolves the infall region well. Detailed chemical models
are therefore very important tools when attempting to interpret the
observations.

4. Distance uncertainties of a factor of two or greater affect the line
asymmetry ratio significantly.

5. Subsonic turbulence which monotonically decreases outwards from the core
centre by a factor of 1.5 produces peak velocity separations in the \hcop\
1--0, 2--1 lines of the order of $\sim$1 km s$^{-1}$. This is comparable to the
peak velocity separations caused by an uncertainty estimate in the distance of
a factor of five. A turbulence which increases with radius by a factor of 1.5
has a smaller effect on the peak velocity separation. Turbulence which
monotonically increases outwards to supersonic values, with a variation of a
factor of three in the core, produces similar peak velocity separations to the
subsonic decreasing case, but results in a much stronger self-absorption in the
lines.

In the future we shall study the sensitivities of line profiles to the infall
dynamics [$v(r)$, in both the collapse and pre-collapse phases] and the
chemical and physical initial conditions. And although this work is of a
general nature, the method of analysis and application can easily be adapted to
model specific sources and instrumental configurations. Concerning the latter,
the next development will be to generate a spectral line interpretation `atlas'
for use in the analysis of {\it Herschel} HIFI datasets.

\vspace{1cm}

\noindent {\bf Acknowledgments}

\noindent{This work was made possible through use of the HiPerSPACE
supercomputer facility at UCL and has made use of the NASA ADS database. We
thank an anonymous referee and D. Ward-Thompson for their critical and in depth
review of this article. We also appreciate helpful comments from N. J. Evans.
YGT acknowledges support from an STFC postdoctoral grant.}

%\section*{References}

\appendix

\section{Model peak line intensities and asymmetry ratios}

\newpage
\setcounter{table}{0}
\begin{table*}
\centering
\begin{minipage}{140mm}
\caption{Temporal evolution of molecular line peak intensities ($I$) and
respective ratios (the line asymmetry ratio) as measured from our simulated
spectra: subscript `b' denotes blue-shifted, `r' denotes red-shifted. The
adopted abundance profiles are discussed in Section 2.1. \tp\ and \tc\ are the
temperature distributions shown in Fig.\,~1.}
\begin{tabular}{lcccc}
\noalign{\vskip3pt} \noalign{\hrule} \noalign{\vskip3pt}
        &$I_{\rm b}$  &$I_{\rm b}$/$I_{\rm r}$     &$I_{\rm b}$ &$I_{\rm b}$/$I_{\rm r}$     \\
        &(K)          &                            &(K)          &                         \\

\noalign{\vskip3pt} \noalign{\hrule}\noalign{\vskip3pt}                                 %x0.074   x0.926
\multicolumn{1}{l}{} &\multicolumn{2}{c}{{\it \tp}} &\multicolumn{2}{c}{{\it \tc}}\\
                                                                                        %H-poor     H-rich
\multicolumn{1}{c}{$t_{\rm coll}$ $=$ 1.36$\times$10$^5$\,yr} &\multicolumn{4}{c}{\hcop}\\
\multicolumn{1}{c}{} &\multicolumn{4}{c}{$n_{\rm mol}$($r$) $=$ `model'} \\
$J$ $=$ 1$\rightarrow$0  &2.619  &0.972         &2.620       &1.025            \\
~~~~~~~~2$\rightarrow$1  &2.910  &0.898         &3.658       &1.088            \\
~~~~~~~~3$\rightarrow$2  &1.136  &0.852         &1.881       &1.166            \\
~~~~~~~~4$\rightarrow$3  &0.390  &0.899         &0.909       &1.316            \\
~~~~~~~~5$\rightarrow$4  & --    &  --          &0.122       &1.369            \\
\multicolumn{1}{c}{} &\multicolumn{4}{c}{$n_{\rm mol}$($r$) $=$ `constant'} \\
$J$ $=$ 1$\rightarrow$0 &1.767      &0.967          &1.897       &1.044                          \\
~~~~~~~~2$\rightarrow$1 &2.234      &0.933          &3.151       &1.178                          \\
~~~~~~~~3$\rightarrow$2 &0.915      &0.944          &1.814       &1.391                          \\
~~~~~~~~4$\rightarrow$3 &0.259      &1.027          &0.901       &1.625                          \\
~~~~~~~~5$\rightarrow$4 &0.027      &--             &0.136       &1.260                          \\
\multicolumn{1}{c}{} &\multicolumn{4}{c}{$n_{\rm mol}$($r$) $=$ `increasing'} \\
$J$ $=$ 1$\rightarrow$0 &1.813                 &0.975                       &1.849               &1.028                                  \\
~~~~~~~~2$\rightarrow$1 &2.130                 &0.920                       &2.852               &1.107                                  \\
~~~~~~~~3$\rightarrow$2 &0.870                 &0.895                       &1.558               &1.244                                  \\
~~~~~~~~4$\rightarrow$3 &0.012                 &--                          &0.721               &1.487                                  \\
~~~~~~~~5$\rightarrow$4 &0.012                 &--                          &0.101               &1.199                                  \\
\multicolumn{1}{c}{} &\multicolumn{4}{c}{$n_{\rm mol}$($r$) $=$ `decreasing'} \\
$J$ $=$ 1$\rightarrow$0 &2.282                &0.959                      &2.443              &1.050                                 \\
~~~~~~~~2$\rightarrow$1 &2.733                &0.913                      &4.091              &1.237                                \\
~~~~~~~~3$\rightarrow$2 &1.278                &0.930                      &2.639              &1.476                                \\
~~~~~~~~4$\rightarrow$3 &0.048                &0.955                      &1.753              &1.646                                \\
~~~~~~~~5$\rightarrow$4 &0.049                & --                        &0.321              &1.734                                \\
\multicolumn{1}{c}{} &\multicolumn{4}{c}{CS}\\
\multicolumn{1}{c}{} &\multicolumn{4}{c}{$n_{\rm mol}$($r$) $=$ `model'} \\
$J$ $=$ 1$\rightarrow$0 &1.509              &0.999                       &1.418              &0.999                                  \\
~~~~~~~~2$\rightarrow$1 &1.201             &0.991                       &1.1541             &0.996                                 \\
~~~~~~~~3$\rightarrow$2  &0.992             &0.993                       &0.989              &1.015                                 \\
~~~~~~~~4$\rightarrow$3  &0.290             &0.999                       &0.300              &1.043                                 \\
~~~~~~~~5$\rightarrow$4  &0.058             &--                          &0.060              &--                                \\
\multicolumn{1}{c}{} &\multicolumn{4}{c}{$n_{\rm mol}$($r$) $=$ `constant'} \\
$J$ $=$ 1$\rightarrow$0 &1.470             &0.996                        &1.435              &1.011                                   \\
~~~~~~~~2$\rightarrow$1 &1.469             &0.973                        &1.580              &1.037                                  \\
~~~~~~~~3$\rightarrow$2 &1.777             &0.991                        &2.280              &1.153                                  \\
~~~~~~~~4$\rightarrow$3 &0.849             &1.062                        &1.368              &1.355                                  \\
~~~~~~~~5$\rightarrow$4 &0.159             &1.108                        &0.377              &1.590                                  \\
\multicolumn{1}{c}{} &\multicolumn{4}{c}{$n_{\rm mol}$($r$) $=$ `increasing'} \\
$J$ $=$ 1$\rightarrow$0 &1.276             &0.997                        &1.220              &0.999                                 \\
~~~~~~~~2$\rightarrow$1 &1.053             &0.988                        &1.052              &1.007                                \\
~~~~~~~~3$\rightarrow$2 &0.993             &0.980                        &1.078              &1.031                                \\
~~~~~~~~4$\rightarrow$3 &0.330             &1.010                        &0.420              &1.128                                \\
~~~~~~~~5$\rightarrow$4 &0.068             &--                           &0.088              &--                                   \\
\multicolumn{1}{c}{} &\multicolumn{4}{c}{$n_{\rm mol}$($r$) $=$ `decreasing'} \\
$J$ $=$ 1$\rightarrow$0 &1.753              &0.996                         &1.708               &1.014                                  \\
~~~~~~~~2$\rightarrow$1 &1.809              &0.964                         &1.968               &1.051                                 \\
~~~~~~~~3$\rightarrow$2 &2.285              &0.974                         &2.943               &1.161                                 \\
~~~~~~~~4$\rightarrow$3 &1.253              &1.042                         &2.051               &1.385                                 \\
~~~~~~~~5$\rightarrow$4 &0.277              &1.158                         &0.697               &1.698                                 \\
\multicolumn{1}{c}{} &\multicolumn{4}{c}{\nthp}\\
\multicolumn{1}{c}{} &\multicolumn{4}{c}{$n_{\rm mol}$($r$) $=$ `model'} \\
$J$ $=$ 1$\rightarrow$0 &0.443             &--                        &0.440              &--                                  \\
%~~~~~~~~2$\rightarrow$1 &1.083             &1.033                        &1.427               &1.190                                 \\
%\multicolumn{1}{c}{} &\multicolumn{4}{c}{p-\amm}\\
%\multicolumn{1}{c}{} &\multicolumn{4}{c}{$n_{\rm mol}$($r$) $=$ `model'} \\
%$J$ $=$ 1$\rightarrow$0 &1.001      & --        &0.918       & --                 \\
%~~~~~~~~2$\rightarrow$1 &0.343      & --         &0.286 & --                           \\

\noalign{\vskip2pt}

\end{tabular}
\end{minipage}
\end{table*}

\setcounter{table}{0}
\begin{table*}
\begin{minipage}{160mm}
\centering \caption{{\it --continued}}
\begin{tabular}{lcccc}

\noalign{\vskip3pt} \noalign{\hrule} \noalign{\vskip3pt}
        &$I_{\rm b}$  &$I_{\rm b}$/$I_{\rm r}$     &$I_{\rm b}$ &$I_{\rm b}$/$I_{\rm r}$    \\
        &(K)          &                            &(K)          &                         \\

\noalign{\vskip3pt} \noalign{\hrule}\noalign{\vskip3pt}
\multicolumn{1}{l}{} &\multicolumn{2}{c}{{\it \tp}} &\multicolumn{2}{c}{{\it \tc}}\\
\multicolumn{1}{c}{$t_{\rm coll}$ $=$ 2.72$\times$10$^5$\,yr} &\multicolumn{4}{c}{\hcop}\\
\multicolumn{1}{c}{} &\multicolumn{4}{c}{$n_{\rm mol}$($r$) $=$ `model'} \\
$J$ $=$ 1$\rightarrow$0 &2.476                 &1.087                         &2.600                 &1.231                                 \\
~~~~~~~~2$\rightarrow$1 &2.394                 &0.984                         &2.857                 &1.207                                \\
~~~~~~~~3$\rightarrow$2 &0.572                 &0.848                         &0.776                 &1.108                                \\
~~~~~~~~4$\rightarrow$3 &0.120                 &1.082                         &0.184                 &1.399                                \\
~~~~~~~~5$\rightarrow$4 &0.008                 &1.000                         &0.012                 &1.024                                \\
\multicolumn{1}{c}{} &\multicolumn{4}{c}{$n_{\rm mol}$($r$) $=$ `constant'} \\
$J$ $=$ 1$\rightarrow$0 &1.718                &0.988                     &1.789             &1.086                                \\
~~~~~~~~2$\rightarrow$1 &1.783                &0.967                     &2.106             &1.174                               \\
~~~~~~~~3$\rightarrow$2 &0.371                &0.917                     &0.513             &1.202                               \\
~~~~~~~~4$\rightarrow$3 &0.064                &1.058                     &0.104             &1.281                               \\
~~~~~~~~5$\rightarrow$4 &  --       &   --                               &  --        &  --               \\
\multicolumn{1}{c}{} &\multicolumn{4}{c}{$n_{\rm mol}$($r$) $=$ `increasing'} \\
$J$ $=$ 1$\rightarrow$0 &1.039                &0.982                   &1.013                 &1.011                                 \\
~~~~~~~~2$\rightarrow$1 &0.769                &0.929                   &0.769                 &1.000                                \\
~~~~~~~~3$\rightarrow$2 &0.115                &1.016                   &0.115                 &1.154                                \\
~~~~~~~~4$\rightarrow$3 &0.025                  &--                      &0.018                 &--                               \\
~~~~~~~~5$\rightarrow$4 &0.002                &--                      & --         & --                                \\
\multicolumn{1}{c}{} &\multicolumn{4}{c}{$n_{\rm mol}$($r$) $=$ `decreasing'} \\
$J$ $=$ 1$\rightarrow$0 &1.877                &1.045                      &1.980                 &1.166                           \\
~~~~~~~~2$\rightarrow$1 &1.935                &1.034                      &2.360                 &1.264                          \\
~~~~~~~~3$\rightarrow$2 &0.411                &0.951                      &0.602                 &1.264                          \\
~~~~~~~~4$\rightarrow$3 &0.073                &1.074                      &0.129                 &1.341                          \\
~~~~~~~~5$\rightarrow$4 &0.005                &--                         &0.009                 &1.007                          \\
\multicolumn{1}{c}{} &\multicolumn{4}{c}{CS}\\
\multicolumn{1}{c}{} &\multicolumn{4}{c}{$n_{\rm mol}$($r$) $=$ `model'} \\
$J$ $=$ 1$\rightarrow$0 &1.261             &1.008                       &1.200              &1.017                             \\
~~~~~~~~2$\rightarrow$1 &0.881             &1.011                       &0.878              &1.063                            \\
~~~~~~~~3$\rightarrow$2 &0.551             &1.041                       &0.575              &1.150                            \\
~~~~~~~~4$\rightarrow$3 &0.116             &1.076                       &0.126              &1.190                            \\
~~~~~~~~5$\rightarrow$4 &0.013             &--                          &0.013              &--                               \\
\multicolumn{1}{c}{} &\multicolumn{4}{c}{$n_{\rm mol}$($r$) $=$ `constant'} \\
$J$ $=$ 1$\rightarrow$0 &1.264             &1.019                      &1.235              &1.048                                 \\
~~~~~~~~2$\rightarrow$1 &1.073             &1.041                      &1.114              &1.126                                \\
~~~~~~~~3$\rightarrow$2 &0.835             &1.095                      &0.945              &1.264                                \\
~~~~~~~~4$\rightarrow$3 &0.208             &1.168                      &0.259              &1.369                                \\
~~~~~~~~5$\rightarrow$4 &0.022             &--                         &0.030              &1.064                                \\
\multicolumn{1}{c}{} &\multicolumn{4}{c}{$n_{\rm mol}$($r$) $=$ `increasing'} \\
$J$ $=$ 1$\rightarrow$0 &1.254             &1.010                       &1.208              &1.025                                  \\
~~~~~~~~2$\rightarrow$1 &0.947             &0.982                       &0.956              &1.040                                 \\
~~~~~~~~3$\rightarrow$2 &0.679             &1.008                       &0.721              &1.128                                 \\
~~~~~~~~4$\rightarrow$3 &0.153             &1.093                       &0.173              &1.240                                 \\
~~~~~~~~5$\rightarrow$4 &0.019             &--                          &0.018              &--                                    \\
\multicolumn{1}{c}{} &\multicolumn{4}{c}{$n_{\rm mol}$($r$) $=$ `decreasing'} \\
$J$ $=$ 1$\rightarrow$0 &1.590              &1.026                      &1.558              &1.066                                  \\
~~~~~~~~2$\rightarrow$1 &1.502              &1.063                      &1.597              &1.186                                 \\
~~~~~~~~3$\rightarrow$2 &1.350              &1.112                      &1.566              &1.307                                 \\
~~~~~~~~4$\rightarrow$3 &0.392              &1.173                      &0.509              &1.436                                 \\
~~~~~~~~5$\rightarrow$4 &0.047              &1.061                      &0.070              &1.193                                 \\
\multicolumn{1}{c}{} &\multicolumn{4}{c}{\nthp}\\
\multicolumn{1}{c}{} &\multicolumn{4}{c}{$n_{\rm mol}$($r$) $=$ `model'} \\
$J$ $=$ 1$\rightarrow$0 &0.180             &--                      & 0.177            &--                                \\
%~~~~~~~~2$\rightarrow$1 &0.452             &1.196                      & 0.527            &1.377                               \\
%\multicolumn{1}{c}{} &\multicolumn{4}{c}{p-\amm}\\
%\multicolumn{1}{c}{} &\multicolumn{4}{c}{$n_{\rm mol}$($r$) $=$ `model'} \\
%$J$ $=$ 1$\rightarrow$0 &0.952      & --             & 0.885      &  --              \\
%~~~~~~~~2$\rightarrow$1 &0.248         &   --        & 0.192        &  --                \\

%\multicolumn{1}{c}{} &\multicolumn{4}{c}{o-\amm}\\
%\multicolumn{1}{c}{} &\multicolumn{4}{c}{$n_{\rm mol}$($r$) $=$ `model'} \\
%$J$ $=$ 1$\rightarrow$0 &0.091      & 0.753       & 0.137      &  0.951              \\
%~~~~~~~~2$\rightarrow$1 &    --     &   --        &  --        &  --                \\

\noalign{\vskip3pt} \noalign{\hrule}\noalign{\vskip3pt}
\end{tabular}
%\begin{description}
%\item[$^a$] `+' indicates that predictions are for the total intensity of the
%multiplet in each case.
%\end{description}
\end{minipage}
\end{table*}

\setcounter{table}{0}
\begin{table*}
\begin{minipage}{160mm}
\centering \caption{{\it --continued}}
\begin{tabular}{lcccc}

\noalign{\vskip3pt} \noalign{\hrule} \noalign{\vskip3pt}
        &$I_{\rm b}$  &$I_{\rm b}$/$I_{\rm r}$     &$I_{\rm b}$ &$I_{\rm b}$/$I_{\rm r}$    \\
        &(K)          &                            &(K)          &                         \\

\noalign{\vskip3pt} \noalign{\hrule}\noalign{\vskip3pt}
\multicolumn{1}{l}{} &\multicolumn{2}{c}{{\it \tp}} &\multicolumn{2}{c}{{\it \tc}}\\
\multicolumn{1}{c}{$t_{\rm coll}$ $=$ 4.08$\times$10$^5$\,yr} &\multicolumn{4}{c}{\hcop}\\
\multicolumn{1}{c}{} &\multicolumn{4}{c}{$n_{\rm mol}$($r$) $=$ `model'} \\
$J$ $=$ 1$\rightarrow$0 &1.809                 &1.447                                        &1.845                            &1.574                                               \\
~~~~~~~~2$\rightarrow$1 &1.336                 &1.001                                        &1.457                            &1.145                                              \\
~~~~~~~~3$\rightarrow$2 &0.227                 &0.986                                        &0.253                            &1.163                                              \\
~~~~~~~~4$\rightarrow$3 &0.038                 &1.028                                        &0.041                            &1.125                                              \\
~~~~~~~~5$\rightarrow$4 &0.002                 &1.000                                        &0.002                            &1.005                                              \\
\multicolumn{1}{c}{} &\multicolumn{4}{c}{$n_{\rm mol}$($r$) $=$ `constant'} \\
$J$ $=$ 1$\rightarrow$0 &1.363                           &1.070                                      &1.360                       &1.144                                            \\
~~~~~~~~2$\rightarrow$1 &0.962                           &0.968                                      &1.009                       &1.090                                           \\
~~~~~~~~3$\rightarrow$2 &0.140                           &1.039                                      &0.151                       &1.217                                           \\
~~~~~~~~4$\rightarrow$3 &0.018                           &1.018                                      &0.019                       &1.044                                           \\
~~~~~~~~5$\rightarrow$4 &  --                            &  --                                       &    --                      &   --                \\
\multicolumn{1}{c}{} &\multicolumn{4}{c}{$n_{\rm mol}$($r$) $=$ `increasing'} \\
$J$ $=$ 1$\rightarrow$0 &0.957                           &0.976                                     &0.902                            &0.981                                               \\
~~~~~~~~2$\rightarrow$1 &0.552                           &0.926                                     &0.503                            &0.943                                              \\
~~~~~~~~3$\rightarrow$2 &0.064                           &1.037                                     &0.053                            &1.062                                              \\
~~~~~~~~4$\rightarrow$3 &0.015                           &--                                        &0.010                            &--                                                 \\
~~~~~~~~5$\rightarrow$4 &0.001                           &--                                        &$<$10$^{-3}$            &    --                                               \\
\multicolumn{1}{c}{} &\multicolumn{4}{c}{$n_{\rm mol}$($r$) $=$ `decreasing'} \\
$J$ $=$ 1$\rightarrow$0 &1.927                           &1.158                                       & 1.957                           &1.265                                               \\
~~~~~~~~2$\rightarrow$1 &1.620                           &1.026                                       & 1.753                           &1.187                                              \\
~~~~~~~~3$\rightarrow$2 &0.270                           &0.951                                       & 0.310                           &1.169                                              \\
~~~~~~~~4$\rightarrow$3 &0.043                           &1.070                                       & 0.049                           &1.170                                              \\
~~~~~~~~5$\rightarrow$4 &0.003                           &1.001                                       & $<$10$^{-3}$                     &1.003                                              \\
\multicolumn{1}{c}{} &\multicolumn{4}{c}{CS}\\
\multicolumn{1}{c}{} &\multicolumn{4}{c}{$n_{\rm mol}$($r$) $=$ `model'} \\
$J$ $=$ 1$\rightarrow$0 &0.951             &1.099                      &0.908              &1.117                              \\
~~~~~~~~2$\rightarrow$1 &0.561             &1.109                      &0.546              &1.160                             \\
~~~~~~~~3$\rightarrow$2 &0.267             &1.148                      &0.261              &1.234                             \\
~~~~~~~~4$\rightarrow$3 &0.042             &1.070                      &0.039              &1.127                             \\
~~~~~~~~5$\rightarrow$4 &0.004             &--                         &0.003              & --                               \\
\multicolumn{1}{c}{} &\multicolumn{4}{c}{$n_{\rm mol}$($r$) $=$ `constant'} \\
$J$ $=$ 1$\rightarrow$0 &1.039             &1.094                       &1.005              & 1.118                                 \\
~~~~~~~~2$\rightarrow$1 &0.739             &1.128                       &0.738              & 1.200                                \\
~~~~~~~~3$\rightarrow$2 &0.408             &1.151                       &0.418              & 1.263                                \\
~~~~~~~~4$\rightarrow$3 &0.075             &1.125                       &0.077              & 1.186                                \\
~~~~~~~~5$\rightarrow$4 &0.006             &1.010                       &0.006              & 1.017                                \\
\multicolumn{1}{c}{} &\multicolumn{4}{c}{$n_{\rm mol}$($r$) $=$ `increasing'} \\
$J$ $=$ 1$\rightarrow$0 & 0.950            &1.043                       &1.005              & 1.118                               \\
~~~~~~~~2$\rightarrow$1 & 0.593            &1.034                       &0.738              & 1.200                              \\
~~~~~~~~3$\rightarrow$2 & 0.298            &1.084                       &0.418              & 1.263                              \\
~~~~~~~~4$\rightarrow$3 & 0.048            &1.086                       &0.077              & 1.186                              \\
~~~~~~~~5$\rightarrow$4 & 0.005            &--                          &0.006              & 1.017                              \\
\multicolumn{1}{c}{} &\multicolumn{4}{c}{$n_{\rm mol}$($r$) $=$ `decreasing'} \\
$J$ $=$ 1$\rightarrow$0 &1.240              & 1.168                       &1.207              &1.209                                   \\
~~~~~~~~2$\rightarrow$1 &0.983              & 1.223                       &0.995              &1.323                                  \\
~~~~~~~~3$\rightarrow$2 &0.589              & 1.179                       &0.618              &1.319                                  \\
~~~~~~~~4$\rightarrow$3 &0.123              & 1.182                       &0.131              &1.293                                  \\
~~~~~~~~5$\rightarrow$4 &0.011              & 1.014                       &0.012              &1.028                                  \\
\multicolumn{1}{c}{} &\multicolumn{4}{c}{\nthp}\\
\multicolumn{1}{c}{} &\multicolumn{4}{c}{$n_{\rm mol}$($r$) $=$ `model'} \\
$J$ $=$ 1$\rightarrow$0 &0.058             & 1.036                      &0.056             & 1.036                              \\
%~~~~~~~~2$\rightarrow$1 &0.164             & 1.217                      &0.170             & 1.313                             \\
%\multicolumn{1}{c}{} &\multicolumn{4}{c}{p-\amm}\\
%\multicolumn{1}{c}{} &\multicolumn{4}{c}{$n_{\rm mol}$($r$) $=$ `model'} \\
%$J$ $=$ 1$\rightarrow$0 &0.581      &  --       & 0.566      &  --               \\
%~~~~~~~~2$\rightarrow$1 &0.090       &     --       & 0.068        &    --               \\
%\multicolumn{1}{c}{} &\multicolumn{4}{c}{o-\amm}\\
%\multicolumn{1}{c}{} &\multicolumn{4}{c}{$n_{\rm mol}$($r$) $=$ `model'} \\
%$J$ $=$ 1$\rightarrow$0 &0.026      &  0.722       & 0.033      &  0.931               \\
%~~~~~~~~2$\rightarrow$1 &   --      &     --       &  --        &    --               \\

\noalign{\vskip3pt} \noalign{\hrule}\noalign{\vskip3pt}
\end{tabular}
%\begin{description}
%\item[$^a$] `+' indicates that predictions are for the total intensity of the
%multiplet in each case.
%\end{description}
\end{minipage}
\end{table*}

\setcounter{table}{0}
\begin{table*}
\begin{minipage}{160mm}
\centering \caption{{\it --continued}}
\begin{tabular}{lcccc}

\noalign{\vskip3pt} \noalign{\hrule} \noalign{\vskip3pt}
        &$I_{\rm b}$  &$I_{\rm b}$/$I_{\rm r}$     &$I_{\rm b}$ &$I_{\rm b}$/$I_{\rm r}$   \\
        &(K)          &                            &(K)          &                         \\

\noalign{\vskip3pt} \noalign{\hrule}\noalign{\vskip3pt}
\multicolumn{1}{l}{} &\multicolumn{2}{c}{{\it \tp}} &\multicolumn{2}{c}{{\it \tc}}\\
\multicolumn{1}{c}{$t_{\rm coll}$ $=$ 5.44$\times$10$^5$\,yr} &\multicolumn{4}{c}{\hcop}\\
\multicolumn{1}{c}{} &\multicolumn{4}{c}{$n_{\rm mol}$($r$) $=$ `model'} \\
$J$ $=$ 1$\rightarrow$0 &0.354                &1.514                         &0.347                 &1.594                             \\
~~~~~~~~2$\rightarrow$1 &0.149                &1.255                         &0.148                 &1.332                            \\
~~~~~~~~3$\rightarrow$2 &0.016                &1.032                         &0.015                 &1.050                            \\
~~~~~~~~4$\rightarrow$3 &0.002                &0.999                         &0.002                 &1.001                            \\
~~~~~~~~5$\rightarrow$4 &  --                 &  --                          &$<$10$^{-3}$          &--                               \\
\multicolumn{1}{c}{} &\multicolumn{4}{c}{$n_{\rm mol}$($r$) $=$ `constant'} \\
$J$ $=$ 1$\rightarrow$0 &0.970                &1.119                          &0.943                 &1.166                            \\
~~~~~~~~2$\rightarrow$1 &0.510                &0.981                          &0.498                 &1.050                           \\
~~~~~~~~3$\rightarrow$2 &0.065                &1.140                          &0.061                 &1.221                           \\
~~~~~~~~4$\rightarrow$3 &0.008                &1.010                          &0.006                 &1.010                          \\
~~~~~~~~5$\rightarrow$4 & --                  &  --                           & --                   &  --                            \\
\multicolumn{1}{c}{} &\multicolumn{4}{c}{$n_{\rm mol}$($r$) $=$ `increasing'} \\
$J$ $=$ 1$\rightarrow$0 &0.160                &1.075                         & 0.150                &1.078                               \\
~~~~~~~~2$\rightarrow$1 &0.050                &1.084                         & 0.045                &1.089                              \\
~~~~~~~~3$\rightarrow$2 &0.006                & --                           & 0.005                & --                                \\
~~~~~~~~4$\rightarrow$3 &0.001                & --                           & $<$10$^{-3}$         & --                                \\
~~~~~~~~5$\rightarrow$4 & --                  & --                           & $<$10$^{-3}$         & --                                \\
\multicolumn{1}{c}{} &\multicolumn{4}{c}{CS}\\
\multicolumn{1}{c}{} &\multicolumn{4}{c}{$n_{\rm mol}$($r$) $=$ `model'} \\
$J$ $=$ 1$\rightarrow$0 & 0.814            & 1.450                     &0.787              & 1.499                              \\
~~~~~~~~2$\rightarrow$1 & 0.551            & 1.418                     &0.543              & 1.500                             \\
~~~~~~~~3$\rightarrow$2 & 0.264            & 1.297                     &0.260              & 1.387                             \\
~~~~~~~~4$\rightarrow$3 & 0.045            & 1.075                     &0.043              & 1.107                             \\
~~~~~~~~5$\rightarrow$4 & 0.004            & 1.008                     &0.003              & 1.009                             \\
\multicolumn{1}{c}{} &\multicolumn{4}{c}{$n_{\rm mol}$($r$) $=$ `constant'} \\
$J$ $=$ 1$\rightarrow$0 & 0.718            & 1.209                    &0.688              & 1.226                            \\
~~~~~~~~2$\rightarrow$1 & 0.410            & 1.201                    &0.397              & 1.249                           \\
~~~~~~~~3$\rightarrow$2 & 0.181            & 1.199                    &0.173              & 1.258                           \\
~~~~~~~~4$\rightarrow$3 & 0.026            & 1.061                    &0.024              & 1.076                           \\
~~~~~~~~5$\rightarrow$4 & 0.002            & 1.007                    &0.002              & 1.009                           \\
\multicolumn{1}{c}{} &\multicolumn{4}{c}{$n_{\rm mol}$($r$) $=$ `increasing'} \\
$J$ $=$ 1$\rightarrow$0 & 0.591            &1.059                      &0.564              &1.060                                \\
~~~~~~~~2$\rightarrow$1 & 0.271            &1.043                      &0.254              &1.053                               \\
~~~~~~~~3$\rightarrow$2 & 0.098            &1.104                      &0.088              &1.117                               \\
~~~~~~~~4$\rightarrow$3 & 0.014            &--                         &0.012              &--                                  \\
~~~~~~~~5$\rightarrow$4 & 0.001            &--                         &0.001              &--                                  \\
\multicolumn{1}{c}{} &\multicolumn{4}{c}{\nthp}\\
\multicolumn{1}{c}{} &\multicolumn{4}{c}{$n_{\rm mol}$($r$) $=$ `model'} \\
$J$ $=$ 1$\rightarrow$0 & 0.005            &1.001                       & 0.005            & 1.003                      \\
%~~~~~~~~2$\rightarrow$1 & 0.013            &1.038                       & 0.012            & 1.048                     \\
%\multicolumn{1}{c}{} &\multicolumn{4}{c}{p-\amm}\\
%\multicolumn{1}{c}{} &\multicolumn{4}{c}{$n_{\rm mol}$($r$) $=$ `model'} \\
%$J$ $=$ 1$\rightarrow$0 &0.140      &  --     & 0.137      & --           \\
%~~~~~~~~2$\rightarrow$1 &0.013     &    --      &0.010    &     --         \\
%\multicolumn{1}{c}{} &\multicolumn{4}{c}{o-\amm}\\
%\multicolumn{1}{c}{} &\multicolumn{4}{c}{$n_{\rm mol}$($r$) $=$ `model'} \\
%$J$ $=$ 1$\rightarrow$0 &0.004      &  0.988     & 0.004      & 1.105           \\
%~~~~~~~~2$\rightarrow$1 &    --     &    --      &      --    &     --         \\

\noalign{\vskip3pt} \noalign{\hrule}\noalign{\vskip3pt}
\end{tabular}
%\begin{description}
%\item[$^a$] `+' indicates that predictions are for the total intensity of the
%multiplet in each case.
%\end{description}
\end{minipage}
\end{table*}

%\end{appendix}

\end{document}